\documentclass[review,12pt]{elsarticle}

\usepackage{amssymb,amsmath}
\usepackage{amsthm}

\usepackage{blkarray}
\usepackage{booktabs}
\usepackage{xargs}   
\usepackage{url,amssymb,bm}
\usepackage[table,pdftex,dvipsnames]{xcolor}
\usepackage[section]{placeins}
\usepackage{diagbox}
\usepackage{multirow}
\usepackage{color, colortbl}
\usepackage{caption}
\usepackage{subcaption}
\usepackage[section]{placeins}

\definecolor{cadmiumgreen}{rgb}{0.0, 0.42, 0.24}

\journal{Renewable Energy}

\begin{document}

\begin{frontmatter}



\title{Measuring Wind Turbine Health \\Using Drifting Concepts}

\author[warsaw]{Agnieszka Jastrzebska\corref{mycorrespondingauthor}}
\address[warsaw]{Faculty of Mathematics and Information Science, Warsaw University of Technology, Poland.}
\cortext[mycorrespondingauthor]{Corresponding author}
\ead{A.Jastrzebska@mini.pw.edu.pl}

\author[HASSELT]{Alejandro Morales Hern\'andez}
\address[HASSELT]{Faculty of Business Economics, Hasselt University, Belgium.}

\author[TILBURG]{Gonzalo N\'apoles}
\address[TILBURG]{Department of Cognitive Science \& Artificial Intelligence, Tilburg University, The Netherlands.}

\author[TALCA]{Yamisleydi Salgueiro}
\address[TALCA]{Department of Computer Sciences, Faculty of Engineering, Universidad de Talca, Campus Curic\'o, Chile.}

\author[HASSELT]{Koen Vanhoof}

\begin{abstract}
Time series processing is an essential aspect of wind turbine health monitoring. Despite the progress in this field, there is still room for new methods to improve modeling quality. In this paper, we propose two new approaches for the analysis of wind turbine health. Both approaches are based on abstract concepts, implemented using fuzzy sets, which summarize and aggregate the underlying raw data. By observing the change in concepts, we infer about the change in the turbine's health. Analyzes are carried out separately for different external conditions (wind speed and temperature). We extract concepts that represent relative low, moderate, and high power production. The first method aims at evaluating the decrease or increase in relatively high and low power production. This task is performed using a regression-like model. The second method evaluates the overall drift of the extracted concepts. Large drift indicates that the power production process undergoes fluctuations in time. Concepts are labeled using linguistic labels, thus equipping our model with improved interpretability features. We applied the proposed approach to process publicly available data describing four wind turbines. The simulation results have shown that the aging process is not homogeneous in all wind turbines.

\end{abstract}




\end{frontmatter}


\section{Introduction}

Operations and maintenance of wind turbines often benefit from machine learning models devoted to helping domain experts make data-driven decisions. In particular, assessing machine performance degradation and predicting component failure allows planning operations and maintenance costs realistically \cite{Jia2018}. This aligns with reliable risk management, which is a~crucial aspect in the renewable energy domain \cite{Avendano2017,Reddy2019}.

Data-driven methods for wind turbine health estimation have gained attention in recent years motivated by several factors. On the one hand, the availability of sensor data to construct such models continues to increase~\cite{Du2020}. On the other hand, developed countries emphasize moving quickly in tackling climate change. In November 2018, the European Union declared that by the year 2050, it should achieve a~climate-neutral status \footnote{\url{https://ec.europa.eu/clima/policies/strategies/2050}}. However, there is an intention to bring closer this deadline\footnote{\url{https://www.elysee.fr/en/emmanuel-macron/2020/10/16/joint-statement-on-the-increase-of-the-eus-2030-climate-target}}, which puts pressure on developing efficient yet human-centric approaches to the management and planning of wind farm operations \cite{Jacobson2018}.

As presented in the literature review drawn up for this study, the prevailing methodology for health index construction is to build a~time series forecasting model and observe the discrepancies between observed and predicted values. Those approaches are model-based since they rely on a secondary model to compute the health index (for example, papers by Zhang et al.~\cite{Zhang2020} and Yang et al.~\cite{Yang2019} follow this scheme). The drawback of such a strategy is that the effectiveness of the model determines the outcome. Moreover, the model fitting stage can be spoiled easily, especially when the data is noisy. Yet another group of the state-of-the-art approaches requires to have recordings concerning several wind turbines and estimates the relative health of the turbines in a data set, like the method of Liu et al.~\cite{Liu2017}. This approach works well only if turbines are of similar mechanical properties and are located in a similar environment, which in practice narrows down its applicability. Moreover, there are methods that operate on a~range of signals from specialized sensors, not installed in all wind turbines. For example, we have methods that require vibration data~\cite{Carroll2019} or pictures of blades \cite{Wu2019} alongside more commonly used data such as signals from the met mast and produced power. Naturally, the applicability of these methods is limited by the data availability. Furthermore, the existing approaches focus on producing one or more numerical scores that evaluate turbine aging. The aspect of the intuitiveness of output presentation is typically left untouched.

To overcome the lack of universal methods for measuring wind turbine efficiency deterioration, in this paper, we present a~conceptually distinct approach based on granular computing. More explicitly, we propose two new methods based on fuzzy concepts to explore aging in wind turbines. Concepts can be understood as abstract information granules that summarize underlying raw data. The first method produces a concept-based description of low and high power production. This description involves a collection of concepts characterizing relative low and relatively high power production. In our approach, we first extract a~collection of concepts in several windows to capture the time flow. Next, we use a concept membership regression model to determine whether there was an increase in low power production and a~decrease in high power production as these would imply aging. The second method measures the drift in all concepts observed in time such that a~large drift indicates (an unwanted) shift in the power production process.

The proposed methods differ significantly from existing approaches since they do not rely on a prediction model but information granules. In contrast to the existing model-based approaches that use prediction accuracy to evaluate the phenomenon of aging (which are qualitatively speaking two essentially different categories), we use granular computing to represent the data and evaluate relative shifts in the data. For our methods to be most informative, we recommend executing performance degradation evaluation separately for several different environmental conditions. In this study, we compute scores separately for different wind and temperature conditions for a given wind turbine to study the relation between the environment and the turbine's performance. Our approach is quite universal since it is suitable to evaluate one or more wind turbines while suing elementary input data found in most physical systems. Moreover, it allows comparing the results for different turbines using both relative and absolute units.

Overall, the rationale of our contribution is to rely on fuzzy concepts to illustrate and estimate the degradation of a wind turbine based on the values of power produced at different operating conditions and granularity levels. Moreover, the fuzzy granulation approach is a suitable formalism to deal with uncertainty in the data. In other words, concepts summarize the information available in raw data and provide a representation robust to noise. As far as we know, the use of fuzzy concepts to measure their drift in time has not been reported in the literature concerning wind turbines.

When it comes to the interpretability, the outputs produced by our models are easy to comprehend by a human being not acquainted well with the applied machine learning methods. In particular, the extracted fuzzy concepts can intuitively be visualized in a two-dimensional scatter plot indicating gradual aging (as quantified through the degradation of the power generation). Furthermore, these information granules are divided into groups and each group has its linguistic label (such as ``low power production'' or ``high power production''). Such linguistic labels allow deriving comprehensible explanations concerning the operation modes.

Aiming at illustrating the reliability of our proposal, we adopt a case study involving four wind turbines. The data is shared publicly by the EDP (\textit{Energias de Portugal}) group under \url{https://opendata.edp.com/}. The results revealed that analyzing wind tower operations performance for various wind speeds and air temperature conditions is very useful. We observe patterns in the scale of deterioration depending on the environmental conditions. Namely, deterioration in power production was most visible for higher wind speeds.
In the analyzed data set, turbines with IDs T07 and T06 deteriorated to a greater extent than turbines T01 and T11.

The remainder of the paper is structured as follows. Section~\ref{sec:literature} discusses relevant health index measures. Section~\ref{sec:method} presents the inner-workings of the new method based on fuzzy information granules. Section~\ref{sec:empirical} addresses a case study focused on the analysis of wind turbine data. This is done through a~case study involving four wind turbines. Section~\ref{sec:conclusion} concludes the paper and provides future research directions.

\section{Literature Review}
\label{sec:literature}

Achieving a~sufficiently high level of profit from generating electric power from wind requires continuous monitoring and systematic maintenance of wind turbines \cite{Avendano2020}. The former task, wind turbine health monitoring, is performed with the use of data coming from a~SCADA (Supervisory Control And Data Acquisition) system \cite{Willis2018,Yang2018}. Such data includes signals from sensors measuring the state of a~wind turbine's components such as gearbox, rotor, generator, or nacelle and data concerning the environment. These indicators are, first and foremost, wind parameters, then the temperature and air pressure. Strategic planning calls for estimating the overall health of a~wind turbine and its deterioration with time \cite{Chen2021}. 

Nowadays, the majority of wind power is generated with large three-bladed horizontal-axis wind turbines \cite{Olauson2018}. In such a system, the main rotor shaft and electrical generator are positioned in the upper part of a~tower, and they must face the wind. While small-scale towers use a~wind vane to position the blades, large-scale towers need a wind sensor and a~yaw system for this purpose \cite{Ouyang2017}. In addition to these components, most wind turbines have a~gearbox, which part takes in changing from a~slow rotation of the blades to a~quicker rotation of the blades and the other way around \cite{Lei2019}. On top of that, we need to mention the power generator located in the so-called nacelle. This rough discussion serves to outline the components ensuring that the turbine is achieving its power production capabilities. Specialized literature on preventive maintenance of wind turbines puts emphasis on monitoring the health of blades, generators, and gearboxes \cite{Stetco2019}.

The literature discusses various approaches to measure wind turbine health, each offering distinct predictive modeling capabilities. Zhang et al.~\cite{Zhang2020} presented a method based on a sliding window approach. In each window, a~third-order polynomial function is fitted to the clean data. Subsequently, the Euclidean distance is used to produce a numerical estimation of turbine health. In contrast, Yang et al.~\cite{Yang2019} proposed a method for component condition monitoring by considering model prediction residuals. Three statistical indexes: Deviation Index, Volatility Index, and Significance Index, were involved in the computations of the health index. Another conceptually similar approach was presented by Zhan et al.~\cite{Zhan2019}, where the authors use Mahalanobis distance between the predicted and real values to obtain the health index. The same kind of an index, comparing predicted and expected values, was discussed in Ren et al.~\cite{Ren2019}. This health index is based on features obtained with a transfer learning approach.

Liu et al.~\cite{Liu2017} stressed the necessity of taking into account wind turbine operating conditions. In the cited work, the authors used kernel density estimation to perform a~preliminary data split. The health index is expressed in a relatively simple manner. They plot all of the data of wind turbines in the group (their method does not work for a single wind turbine) in a~single wind speed versus power chart. If a given wind turbine's data points deviate from the majority of data from all turbines, they assume it is at a~high risk of failure. This method is devoted to wind farms with machines of the same properties located nearby. The method becomes ineffective when this assumption is not fulfilled. Zhang et al.~\cite{Zhang2019} suggest distinguishing four operational condition parameters, which are determined in an unsupervised manner. By using the historical data concerning expected turbine operation, a health benchmark model was constructed. The final health index was obtained using the Mahalanobis distance. Other studies also focus on analyzing the operating conditions. For example, Tewolde et al.~\cite{Tewolde2017} discussed a~method dedicated to offshore wind turbines. 

A lively discussion concerns the variables to be used to monitor the turbine. For example, Koukoura~\cite{Koukoura2018} mainly used components' temperature to analyze turbine health, while Carroll et al.~\cite{Carroll2019} proposed to include vibration data (not that popular elsewhere) among data from other sensors. Ren et al.~\cite{Ren2021} use vibration data exclusively. Li et al.~\cite{Li2019} advocated for using various SCADA parameters but with an adaptive weight fitted using the analytic hierarchy process. Moreover, there exist approaches that use imaging to analyze wind turbine operations, as presented by Wu et al.~\cite{Wu2019}. Admittedly, many approaches (e.g., Song et al.~\cite{Song2018}) employ a few variables, for instance, wind speed, generated power, and generator speed. 

An interesting analysis of logs from three wind turbines that have gearboxes in different damage stages was delivered by Lopez et al.~\cite{Lopez2019}. The authors proposed a correlation-based measure to evaluate turbine health. In contrast, Tcherniak~\cite{Tcherniak2016} focused on rotor health and proposed a method for analyzing structural damages of rotor blades.

Another open problem in the wind turbine health analysis is the lack of methods that focus on user experience. 
The existing approaches concentrate on producing a numerical estimation of wind turbine health, neglecting the interpretation of the produced score \cite{Zhao2018}. In this paper, we refocus the approach to health index construction using fuzzy information granules. The proposed methods use linguistic labels and intuitive visualizations that complement a numerical health index. As a result, the outputs produced by our methods are straightforward to interpret by domain experts who are not experts in machine learning. 

\section{Measuring the Health of Wind Turbines with Fuzzy Drifting Concepts: Two Approaches}
\label{sec:method}

In this section, we outline the details of the new approach to wind turbine health evaluation. The approach covers two health indexes measuring power production deterioration through changes observed in concepts. The approach as a whole is composed of both visual and quantitative elements. The method consists of the following steps: (i) Data pre-processing, (ii) Operating conditions binning, (iii) Concepts extraction, and (iv) Concept analysis and computation of the quantitative indexes (two separate methods). The first method extracts information about relative low and relative high power production and then evaluates whether there was an increase in the former and decrease in the latter that would suggest aging. The second approach measures the discrepancy in power production in time by comparing concepts that denote extreme power production conditions. 

The primary information concerning wind turbine health is the amount of power generated under certain environmental conditions. Therefore, the procedure evaluates the extent to which the amount of produced power decreased with time. It is worth mentioning that we will build separate concept-based models for specific wind and temperature ranges. In the next subsections, we will explain the rationale of each step in detail.

\subsection{Data Preprocessing} 
The first step relies on elementary data cleaning to produce active power time series. More explicitly, we remove observations corresponding to very low and very high wind values. To determine what should be the cutting thresholds for low and high wind values, we can use, for example, the turbine's theoretical power curve if available. If we do not wish to use a theoretical power curve, we can inspect the wind-power plot. In this plot, we ought to determine wind values between which we have the diagonal part of the data scatter. Narrowing down the data in this way lets us avoid taking into account instances describing the take-off and the saturation of the wind turbine. In our case, we analyzed power generated for wind values between $4.5$ and $9$ m/s, but this may be tuned for each data set by looking at the wind-power curve. Equation \eqref{eqn:ratiopower} shows how to compute the ratio between power value and wind value for each moment in time,

\begin{equation}
    pw_i = \frac{p_i}{w_i}, i=1, 2, \ldots, L
\label{eqn:ratiopower}
\end{equation}

\noindent where $L$ denotes the total number of observations in the data set, $p_i$ corresponds to the power value in time $i$ and $w_i$ corresponds to the wind value in time $i$. In our approach, we use the ratio from Equation \eqref{eqn:ratiopower} to determine which observations fall between the first and the third quantile of data distribution. Data points that fall into this range are kept, while remaining data points are assumed to be outliers and are removed.

\subsection{Operating Conditions Binning}

The second step of the procedure aims at splitting the cleaned data into segments (bins), each sub-bin describing different environmental conditions. We recommend using two time series: wind speed and air temperature. If the air temperature measurements are not available, the we could use the air pressure variable as an alternative.

Wind speed will be analyzed in bins each of an interval length of 0.5~m/s, that is $4.5, 5.0, 5.5, \ldots$ and so on. The length of the interval equal to $0.5$ is frequently seen in the literature, cf.~\cite{Dorrego2021}.  Temperature will be split into intervals determined using a~clustering procedure (as it is performed in other studies, including \cite{Zhang2019}). In our study, we considered four clusters for temperature values. The $k$-means algorithm \cite{Kanungo2002} is a~suitable choice to determine the segments. Next, we use the obtained centroids to define the limits of intervals to distinguish different conditions. Therefore, a~new instance (that is to say, a~given temperature value) will be accounted to the cluster whose centroid is the closest to it.

\begin{figure}[!h]
    \centering
    \hspace{0pt}\includegraphics[width=0.75\textwidth]{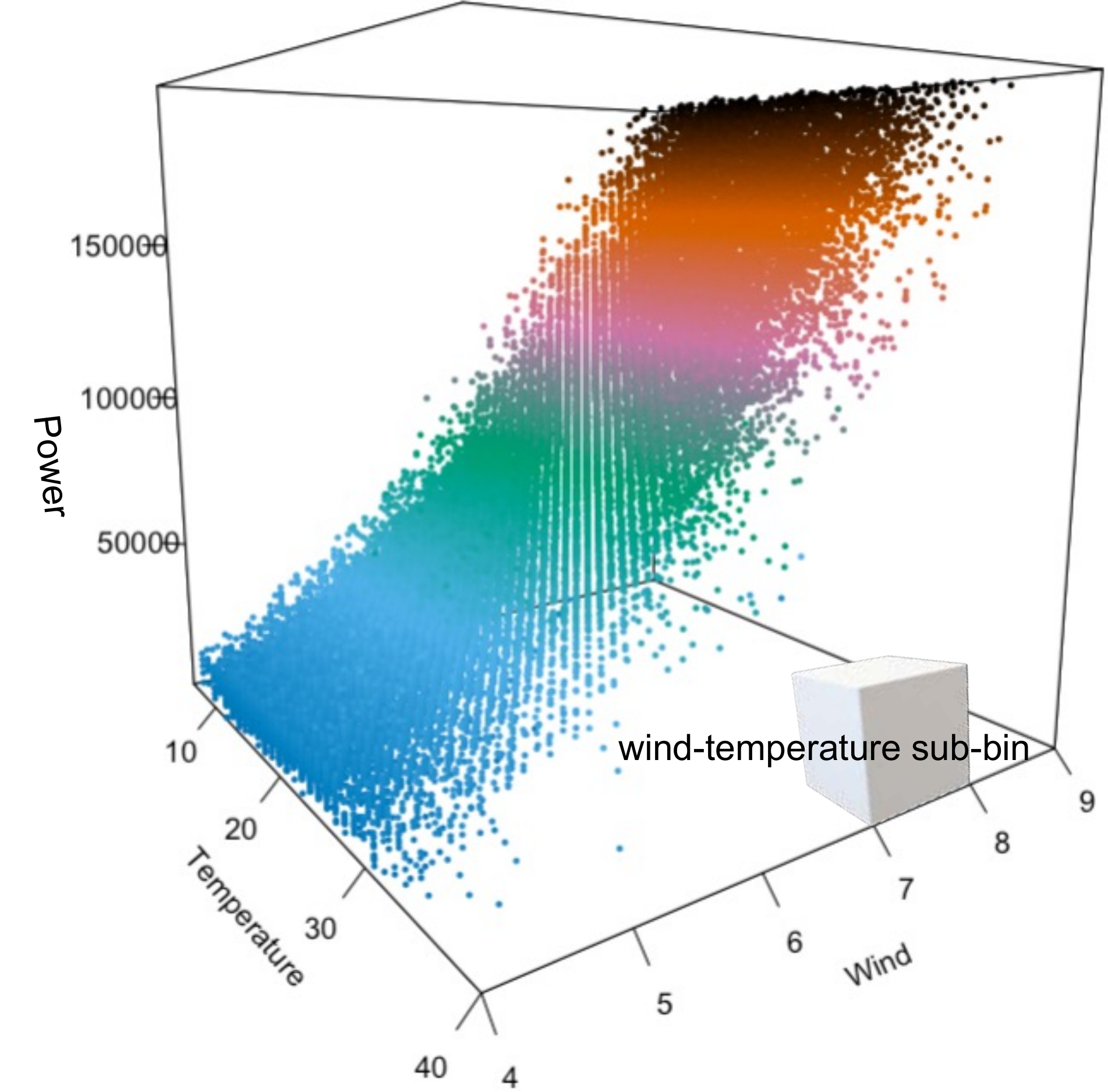} 
    \caption{Granulation of the feature space defined by relevant variable where a~sub-bin is visualized as a~small hyper-cube. In this example, the sub-bin is empty because the data contains no points in the region covered by the sub-bin.}
    \label{fig:cube}
\end{figure}

Figure \ref{fig:cube} illustrates the result of partitioning the observations concerning selected wind and temperature conditions. The reader can notice that this granular approach builds hyper-cubes within the high-dimensional space defined by the temperature, wind and power variables.

\subsection{Concepts Extraction}

In this step, we measure wind turbine health for each bin (that is to say, for a specific wind value range and air temperature range). By doing that, we order the data according to the timestamp and split it into $R$~windows, where $R$~is a parameter denoting the window length. For example, if the data is collected with a~10-minutes frequency, then $R$=40 or $R$=60 would be reasonable configurations. Each window $w$ contains $P$ observations arranged in a~sequence of active power values as follows:

\begin{equation}
    z_{1}^{(w)}, z_{2}^{(w)}, z_{3}^{(w)}, \ldots, z_{P}^{(w)}.
\label{eqn:scalar_time_series}
\end{equation}

The superscript $(w)$ was placed to emphasize that Equation \eqref{eqn:scalar_time_series} concerns a window, not the entire time series. Next, we transform the active power sequence in Equation \eqref{eqn:scalar_time_series} into a~two-dimensional space by computing successive increments of time series values. The $i$-th data point is represented with a~pair $(z_{i}^{(w)}, dz_{i}^{(w)})$ denoting the $i$-th amplitude and the $i$-th change (increment) of amplitude, where $dz_{i}^{(w)}=z_{i}^{(w)}-z_{i-1}^{(w)}$. This means that the active power can be represented with the a~sequence:

\begin{equation}
    (z_{2}^{(w)},dz_{2}^{(w)}),(z_{3}^{(w)},dz_{3}^{(w)}),\ldots,(z_{P}^{(w)},dz_{P}^{(w)}).
\label{eqn:dynamics2_time_series}
\end{equation}

The start index is two because we do not have a~change of amplitude to be set together with the first observation. 

For each window (now represented in two-dimensional space), we extract concepts using fuzzy $c$-means clustering \cite{Bezdek1981}. This algorithm extracts centroids representing the discovered clusters while being endowed with a membership function that links data points to each centroid. Fuzzy $c$-means was pproposed by J.~Dunn in 1973  \cite{Dunn1973} and improved by J.~Bezdek in 1981  \cite{Bezdek1981}. The number of centroids, denoted as $C$, has to be set a~priori. The clusters are fuzzy since one data point belongs to more than one cluster at the same time with a membership value in the $[0,1]$ interval. Moreover, the sum of all membership values for a~single point adds up to 1.

In the proposed approach, we recommend setting $C$=3 such that we can obtain centroids representing \textit{low}, \textit{moderate}, and \textit{high} power values. Wind turbine wear will manifest with: (i) Memberships to centroids representing high values that diminish with time, and (ii) memberships to centroids representing low values that increase with time.

The fuzzy c-means algorithm creates a~matrix $\mathbf{U} = [\mu_{ij}], \mu_{ij} \in [0,1], j=1, \ldots, C, i=1, \ldots, P-1$. Each element $\mu_{ij}$ denotes the degree of membership of the $i$-th data point to the $j$-th fuzzy cluster. 
The procedure involves a~fuzzification coefficient $m \in \mathbb{R}$, such that $1.0 < m < \infty$. Large $m$ attenuates memberships values, thus, we achieve less differentiation in concept belongingness, or in other words, clusters are fuzzier. In contrast, $m$ close to 1 makes the fuzzy c-means act as a crisp clustering algorithm. In \cite{HathawayBezdek2001}, the authors recommend setting $m$ to $2.0$.

The fuzzy $c$-means algorithm is iterative by nature, which involves minimizing the objective function in each iteration:

\begin{equation}
    J_m = \sum_{i=1}^{{P-1}} \sum_{j=1}^C \mu_{ij}^m\Vert\mathbf{z}_i - \mathbf{v}_j\Vert^2,
\label{eqn:fcmeans_objectivefun}
\end{equation}

\noindent where $\mathbf{z}_i$ is the $i$-th element of clustered data, $\mathbf{v}_j$ is a two-dimensional vector with the coordinates of the $j$-th cluster, $m$ is the fuzzification coefficient, $\mu_{ij}$ is the degree of membership of $\mathbf{z}_i$ to the \mbox{$j$-th} cluster, $\Vert\cdot\Vert$ is any norm to evaluate the similarity between a data point  and a centroid \cite{Zhang2018}. 

The iterative procedure adjusts the membership values $\mu_{ij}$ from the matrix $\mathbf{U}$ and the cluster centers $\mathbf{v}_j$ as follows:

\begin{equation}
    \mu_{ij} = \frac{1}{\sum\limits_{k=1}^C \Big( \frac{\Vert \mathbf{z}_i - \mathbf{v}_j \Vert}{\Vert \mathbf{z}_i - \mathbf{v}_k \Vert} \Big)^{2/(m-1)} }
\label{eqn:fcmeans_memfun}
\end{equation}

and 

\begin{equation}
    \mathbf{v}_j = \frac{\sum\limits_{i=1}^{P-1} \mu_{ij}^m \cdot \mathbf{z}_i}{\sum\limits_{i=1}^{P-1} \mu_{ij}^m}.
\label{eqn:fcmeans_cj}
\end{equation} 

The procedure can be terminated when the greatest change in $\mathbf{U}$ from $k$-th to $(k+1)$-st iteration has been lower than a given $\varepsilon$ threshold or when a~predefined number of iterations has been exceeded \cite{Pimentel2018}.

As mentioned, concept extraction is performed on each window. Next, we sort the concepts in decreasing order (high, moderate, low). After the completion of this step, we obtain $R \cdot C$ concepts labeled with the window number for which they were extracted.
 
\subsection{Concept Analysis and Health Index Approaches}

Extracted concepts can be used to visualize and evaluate the degradation tendency in the data. Overall, we propose two strategies to quantify the power production deterioration with time: (i) A regression-based method utilizing drifting concept memberships, and (ii) A distance-index-based method that measures the degree of concept drift.


Both approaches can be utilized to a data set of any number of wind turbines (as we mentioned in the literature review, some methods require the data set to cover several wind turbines). The advantage of the regression-based method is that it is more detailed. It focuses on measuring the decrease or increase in power production in a certain range. The ranges are described with linguistic labels (``low power production'', ``high power production''). The outputs produced by this model are sentences with the form: ``We observed an increase/decrease in low/high power production''. Moreover, we can give a precise evaluation of the increase/decrease observed in power production to quantify the turbine's aging. The second health index approach is more compact. It characterizes the turbine's health with a single value taking into consideration different conditions.

\subsection{Computation of the Regression-Based Health Index}

The first procedure is performed using a regression analysis involving the membership values of observations to the clusters representing the highest time series values. Let us recall that concepts reside is a two-dimensional space of produced power value and change of value (see Equation \eqref{eqn:dynamics2_time_series}) and that each concept is represented with a centroid. 
Notice that centroids are two-dimensional points located in the value and change of value space. We sorted the centroids according to the power value (the first dimension) in decreasing order: the first centroid represents data points with high values of generated power while the third centroid represents underlying data points with low values of generated power.

Let us assume that we process the $k$th window (out of $R$ windows) and obtain a series with the form $\mu_{1_k}^{(1_k)}, \mu_{2_k}^{(1_k)}, \ldots, \mu_{P-1_k}^{(1_k)}$, where $\mu_{i_k}^{(1_k)}$ denotes the membership of the $i$-th data point from the $k$-th window to the first centroid computed for this window. $P-1$ is the number of data points in the two-dimensional space of value and change of value. The ``minus 1'' is because we computed the lags, as in Equation \eqref{eqn:dynamics2_time_series}. The first centroid in the $k$th window is denoted as $(1_k)$ in the superscript. We concatenate the series from each window to obtain the following ordered sequence:
\vspace{-3pt}
\begin{equation}
    \mu_{1_1}^{(1_1)}, \mu_{2_1}^{(1_1)}, \ldots, \mu_{P-1_1}^{(1_1)}, \mu_{1_2}^{(1_2)}, \mu_{2_2}^{(1_2)}, \ldots, \mu_{P-1_2}^{(1_2)}, \ldots, \mu_{1_R}^{(1_R)}, \mu_{2_R}^{(1_R)}, \ldots, \mu_{P-1_R}^{(1_R)}.
\label{eqn:regression_pre}
\end{equation}

For the clarity of notation, in Equation \eqref{eqn:regression_pre}, we retain the original order of values while replacing the nested indexing with a single-level indexing. The following equation formalizes this compact notation:

\begin{equation}
    \mu_{1}^{(1)}, \mu_{2}^{(1)}, \ldots, \mu_{N}^{(1)}.
\label{eqn:regression_post}
\end{equation}

\noindent where $\mu_{1}^{(1)}$ corresponds to $\mu_{1_1}^{(1_1)}$ and so on, whereas $N = (P-1) \cdot R$.
Next, we compute a regression model in the form:

\begin{equation}
    {\mu}_{i}^{(1)} = a^{(1)} x_i + b^{(1)} + \varepsilon_i^{(1)}.
\label{eqn:regression_mu}
\end{equation}

The procedure leading to the regression model in Equation \eqref{eqn:regression_mu} is computed for the specified wind speed and temperature sub-bin and shall be repeated for each sub-bin. In this regression model, which is fitted with the data given in Equation \eqref{eqn:regression_post}, $x$ denotes the observation order in time. A negative value of the slope $a^{(1)}$ indicates aging since it suggests a decrease in the membership to the centroids corresponding to the high power production for the given wind and temperature conditions. We assume that a linear model will be a simple, yet suitable, to model the turbine aging.

An analogous analysis can be performed for the concepts representing the low time series values. According to our method, aging would be manifested as an increasing membership to these fuzzy concepts. Aiming at evaluating whether the membership values increase or decrease with time, we  can compute a regression model with the following form:

\begin{equation}
    {\mu}_{i}^{(3)} = a^{(3)} x_i + b^{(3)} + \varepsilon_i^{(3)}
\label{eqn:regression_mu3}
\end{equation}

\noindent where the superscript $(3)$ indicates that we deal with the third (and last) concept concerning low power production. The regression model given by Equation \eqref{eqn:regression_mu3} is fitted using the sequence below:

\begin{equation}
    \mu_{1}^{(3)}, \mu_{2}^{(3)}, \ldots, \mu_{N}^{(3)}.
\label{eqn:regression_post3}
\end{equation}

An important feature of this aging index is that coefficients $a^{(1)}$ and $a^{(3)}$ are comparable for any wind turbine in any data set. Membership values are always in $[0,1]$, thus the models are created on normalized data. These values describe relative aging, not in natural units (like watt-hours) that would always depend on mechanical properties.

\subsection{Computation of the Centroid Distance-Based Health Index}

The idea behind the second health index is to measure the discrepancy between \textit{high} and \textit{low} power production centroids (concepts). Let us recall that for each window (we have $R$ windows ordered in time), we created $C$ clusters to represent the values falling into it. That is, we have $C$ types of clusters that describe low, moderate, and high power production. Let us recall that we marked them with different symbols in the plots (triangles, circles, and squares). The total number of centroids describing high power production is $R$, the same for moderate, and low production. Visual interpretation of $R \cdot C$ centroids may be inconvenient. Thus, we propose an additional aggregation option to simplify the interpretation.

Let us formalize this discussion by denoting the extracted centroids as presented in Equations \eqref{eqn:vs_high}, \eqref{eqn:vs_med}, and \eqref{eqn:vs_small}.
\vspace{-5pt}
\begin{equation}
    \textbf{v}_1^{(1)}, \textbf{v}_2^{(1)}, \ldots, \textbf{v}_R^{(1)}
\label{eqn:vs_high}\vspace{-4pt}
\end{equation}
\begin{equation}
    \textbf{v}_1^{(2)}, \textbf{v}_2^{(2)}, \ldots, \textbf{v}_R^{(2)}
\label{eqn:vs_med}\vspace{-2pt}
\end{equation}
\begin{equation}
    \textbf{v}_1^{(3)}, \textbf{v}_2^{(3)}, \ldots, \textbf{v}_R^{(3)}
\label{eqn:vs_small}
\end{equation}

In Equation \eqref{eqn:vs_high}, are the centroids representing high power production.
Equation \eqref{eqn:vs_med} represents moderate power production.
In Equation \eqref{eqn:vs_small}, are the centroids concerning low power production.
They are ordered according to the window for which they were created.

We can generalize further the information carried by these $R \cdot C$ centroids by splitting each type of centroids into two clusters, again using a~centroid-based algorithm. In particular, we again use the fuzzy c-means method (for consistency).
That is, we forward the sequence from Equation \eqref{eqn:vs_high} to the input of the fuzzy c-means clustering algorithm. We forward the sequence from Equation \eqref{eqn:vs_med} and run the fuzzy c-means for this data. Finally, we take the sequence given in Equation \eqref{eqn:vs_small} and pass it to the input to the fuzzy c-means algorithm. Each time we set the desired number of produced clusters to two. In other words, we obtain a~pair of clusters in each case: low, moderate, and high power production. Let us mention that the input data (Equation \eqref{eqn:vs_high} and Equation \eqref{eqn:vs_small}) are two-dimensional vectors because we operate in the space of value and change of value.

Let us present and compare the interpretation of the original centroids, listed in Equation \eqref{eqn:vs_high} -- Equation \eqref{eqn:vs_small} and the new centroids, obtained after the fuzzy c-means was run for the old centroids. The centroids obtained in the first stage correspond to different moments in time (because they were computed for windowed data). Let us recall that in the plots, we used colors to account for the time flow. The new centroids do not capture time flow, but they can be used to describe the discrepancy in the data. 
At the present stage, we extract two centroids for each group. We can sort them and interpret one as a~centroid describing low values and the other one as a~centroid describing high power production values. Obtained pairs of centroids are noted in Equations \eqref{eqn:vs_high_second}, \eqref{eqn:vs_med_second}, and \eqref{eqn:vs_small_second}.
\vspace{-3pt}
\begin{equation}
    \textbf{v}_L^{(1)}, \textbf{v}_H^{(1)}
\label{eqn:vs_high_second}
\end{equation}
\begin{equation}
    \textbf{v}_L^{(2)}, \textbf{v}_H^{(2)}
\label{eqn:vs_med_second}
\end{equation}
\begin{equation}
    \textbf{v}_L^{(3)}, \textbf{v}_H^{(3)}
\label{eqn:vs_small_second}
\end{equation}

In each pair, the first centroid corresponds to low power production values (the letter $L$ in the subscript). The second centroid corresponds to high power production values (with the letter $H$ in the subscript). The intuition of this measure is that the more separated the centroids, the more aging in the turbine. This distance will be zero in an ideal world scenario where there is no change in the performance. Equation \eqref{eqn:dist_index} shows the distance index ($DI$) formalizing this turbine health index: 
\vspace{-3pt}
\begin{equation}
    DI = \sum_{i=1}^{3} d(\textbf{v}_{L}^{(i)},\textbf{v}^{(i)}_{H}).
\label{eqn:dist_index}
\end{equation}

\noindent where $d(\cdot,\cdot)$ denotes any distance measure (e.g., the Euclidean distance). We interpret large $DI$ values as a negative phenomenon. Let us recall that the centroids are extracted for fixed wind and air temperature values, so we are justified to demand as small differentiation in machine performance as possible. Notice that the $DI$ health index will provide better insight when computed for more than one wind-temperature sub-bin.

\section{Case Study -- EDP Data Set}
\label{sec:empirical}
\vspace{-3pt}
In this section, we present the application of the proposed procedure to EDP (\textit{Energias de Portugal}) data set, which is publicly available \url{https://opendata.edp.com/}. This case study concerns four wind turbines SCADA signals for the years 2016 and 2017.
Further parts of this section are split into three subsections. In Subsection~\ref{subsec:detailed_empirical}, we present an introductory analysis of the results concerning the turbine with ID T11.
We chose to discuss T11 in greater detail because later comparative analysis showed that the deterioration of performance of this turbine was relatively moderate. We did not want to analyze first the most deteriorated or the least deteriorated one.
Subsection~\ref{subsec:comparative_empirical} compares the results for four turbines in the EDP data set and highlights interesting observations. Subsection \ref{subsec:addition} presents the application of the Distance Index for turbine performance evaluation. 

\begin{figure}
    \centering
    \begin{subfigure}{1\textwidth}
        \centering
        \includegraphics[width=.32\textwidth]{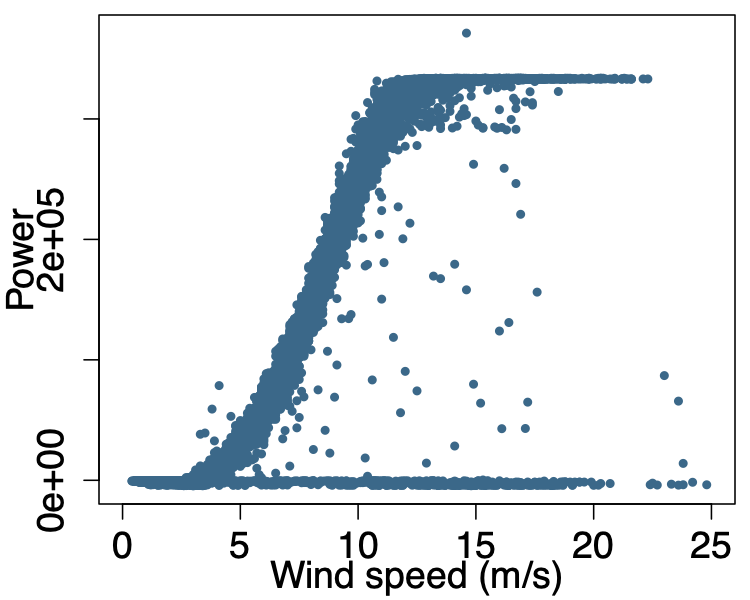}
        \includegraphics[width=.32\textwidth]{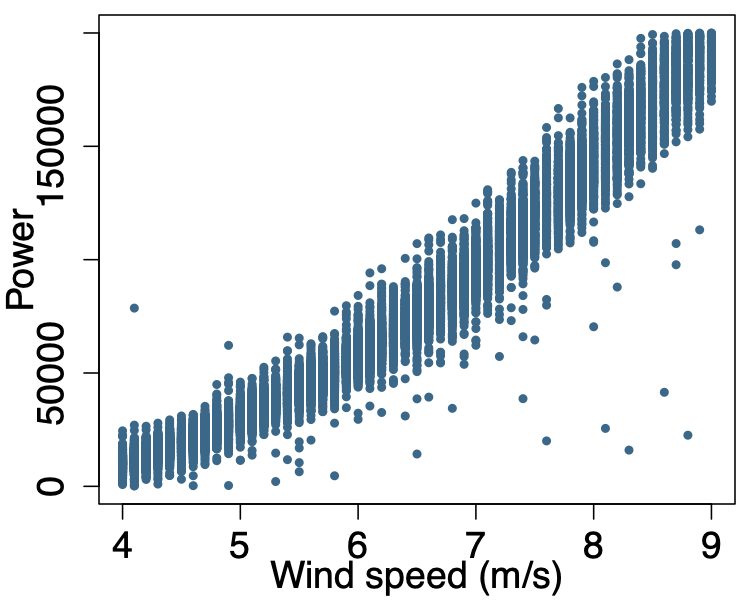}
        \includegraphics[width=.32\textwidth]{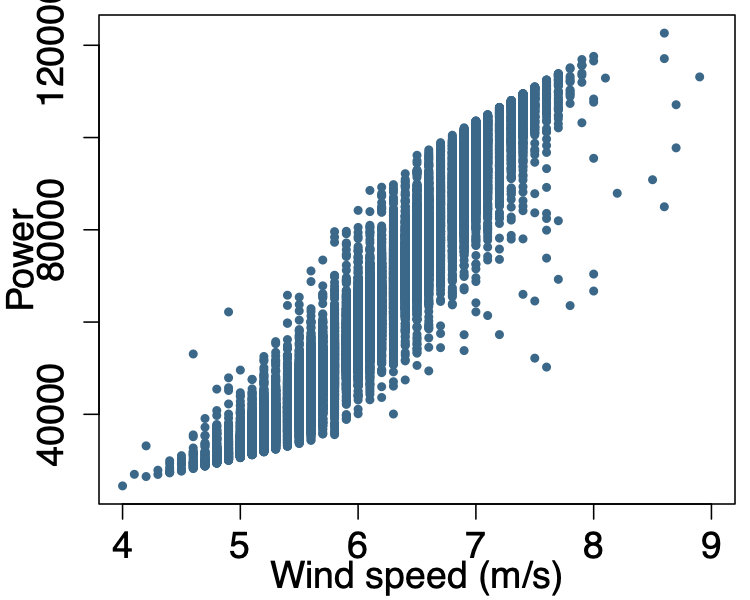}
        \caption{T01}
    \end{subfigure}
    \hfill
    \begin{subfigure}{1\textwidth}
        \centering
        \includegraphics[width=.32\textwidth]{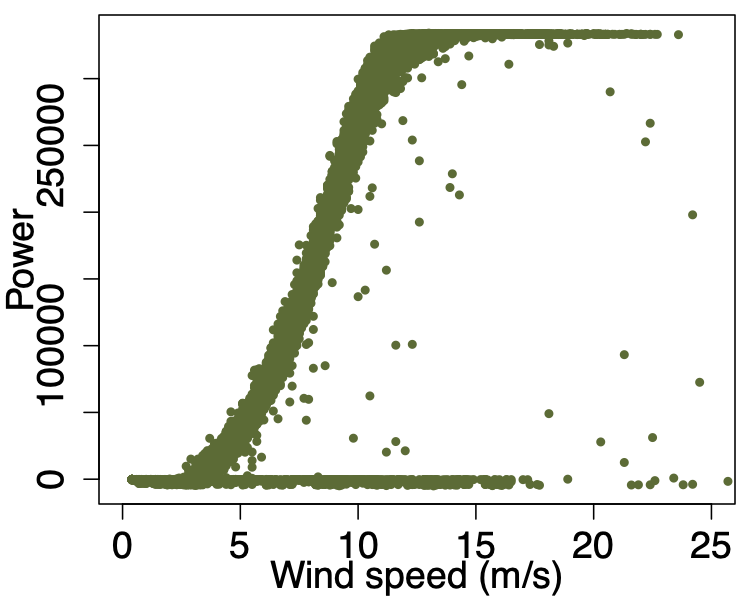}
        \includegraphics[width=.32\textwidth]{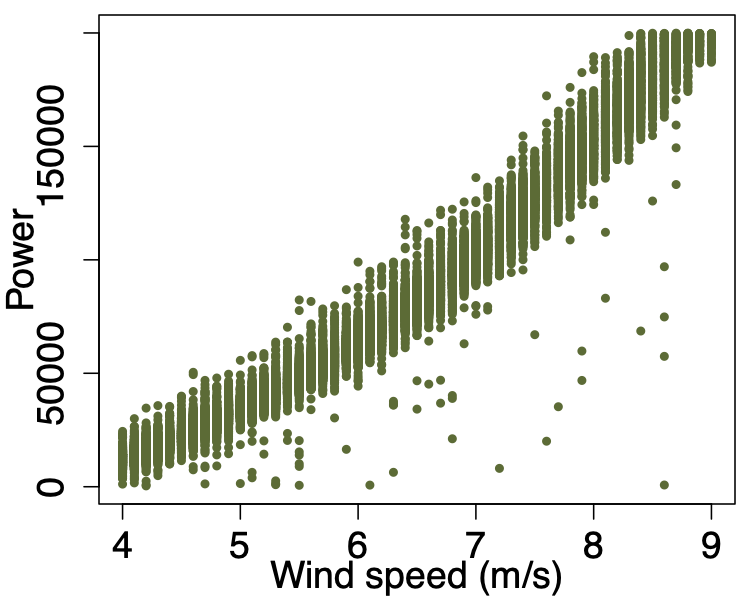}
        \includegraphics[width=.32\textwidth]{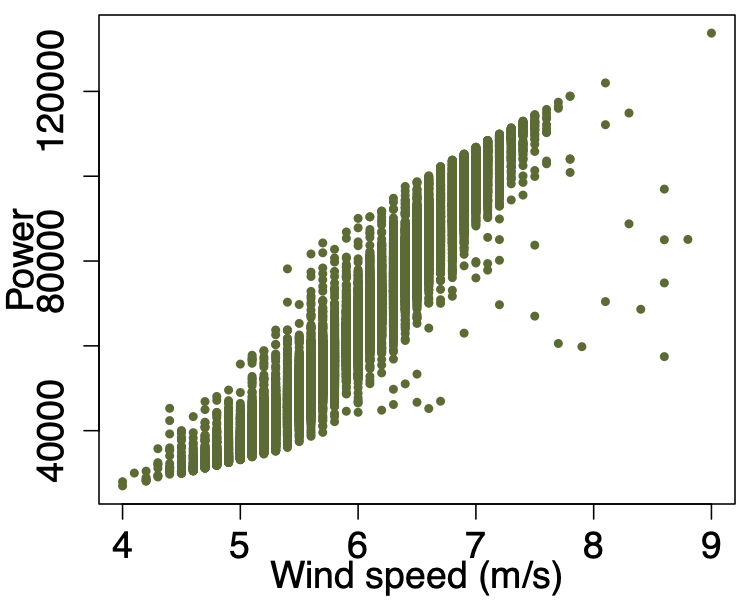}
        \caption{T06}
    \end{subfigure}
    \hfill
    \begin{subfigure}{1\textwidth}
        \centering
        \includegraphics[width=.32\textwidth]{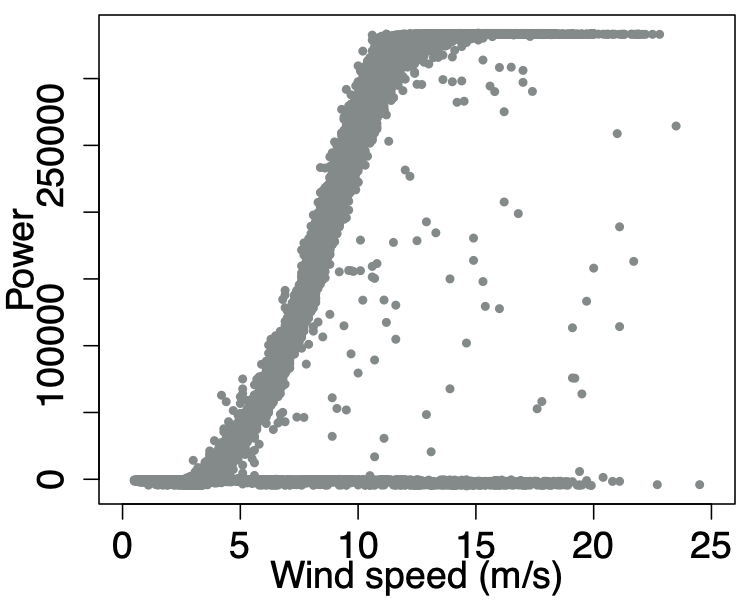}
        \includegraphics[width=.32\textwidth]{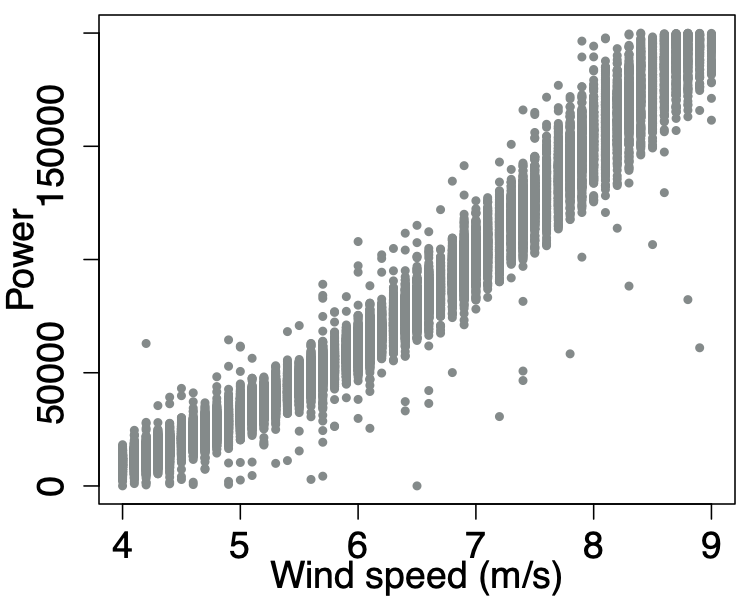}
        \includegraphics[width=.32\textwidth]{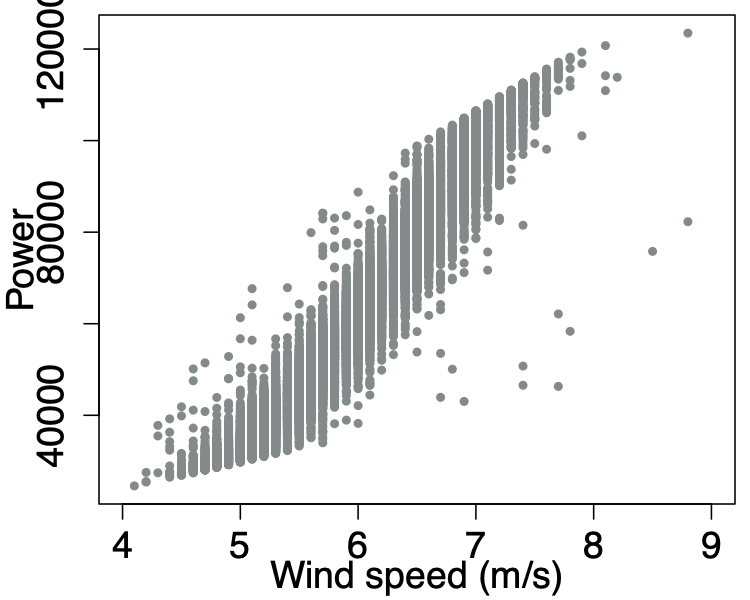}
        \caption{T07}
    \end{subfigure}
    \hfill
    \begin{subfigure}{1\textwidth}
        \centering
        \includegraphics[width=.32\textwidth]{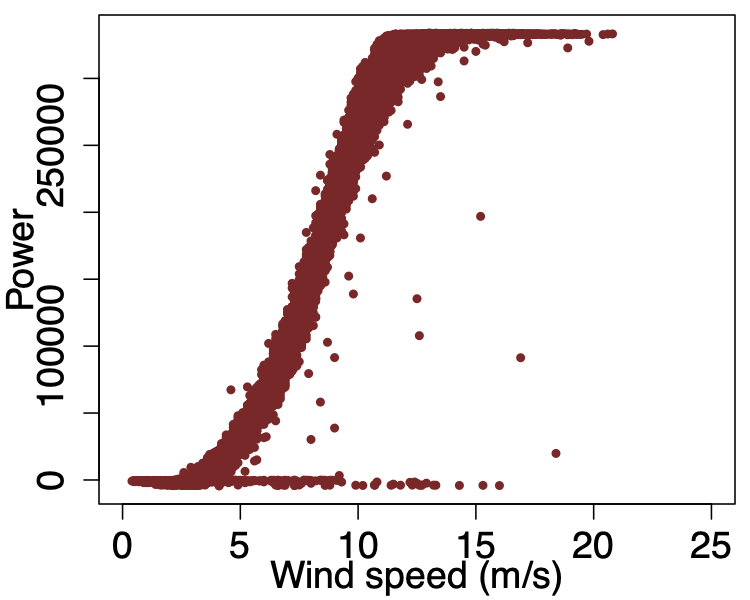}
        \includegraphics[width=.32\textwidth]{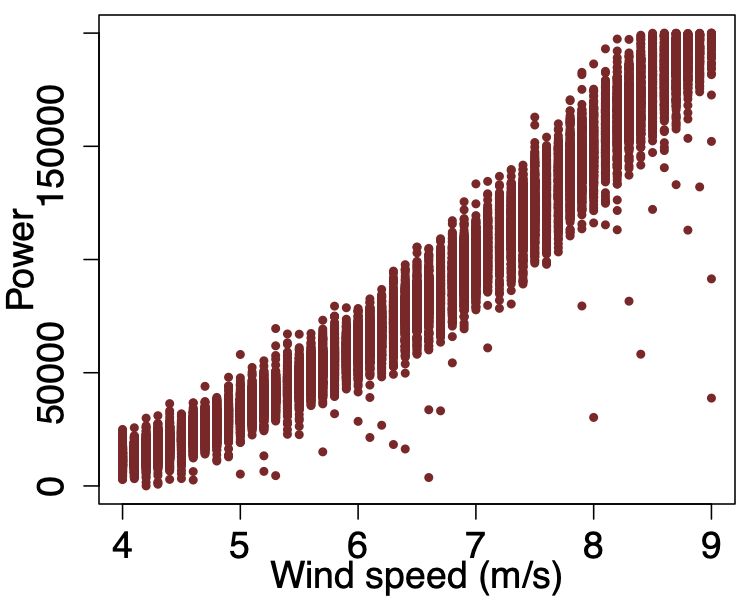}
        \includegraphics[width=.32\textwidth]{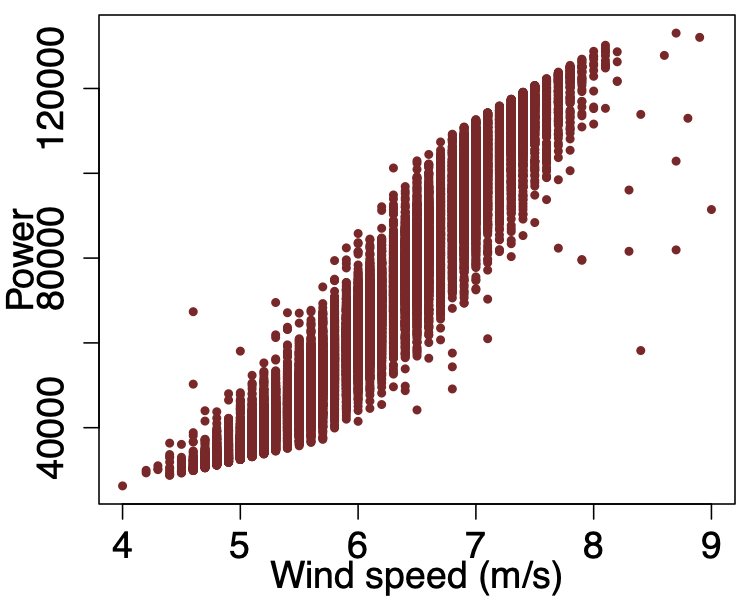}                  
        \caption{T11}
    \end{subfigure}
    \caption{Power versus wind speed for each wind turbine (turbine IDs are given in sub-captions). The first plots contain all data points available in the EDP data sets. The second plots contain data points after narrowing down the wind range to $[4, 9)$ m/s. Third plots show the remaining points after cleaning the extreme values.}
    \label{fig:power_curve}
\end{figure}

Figure \ref{fig:power_curve} shows some statistics concerning the power and wind speed variables of all wind turbines. More explicitly, the first column displays plots of the relationship between power and wind speed for each analyzed wind turbine. Data scatter follows the usual sigmoid-like shape. The data is noisy as expected (i.e., we have plenty of outliers and in T01, and there is one anomalous observation). The second column concerns the data after narrowing it down to the wind speed values between 4 and 9 m/s. The third column contains plots after selecting observations between the first and third quantile with the use of the ratio given in Equation \eqref{eqn:ratiopower}.

After cleaning, the data is transformed to the two-dimensional space of value and change of value as shown in Equation \eqref{eqn:dynamics2_time_series}. The available data points are partitioned into bins, each bin corresponding to a~specific range of wind values. In this case study, bins started at 5~m/s and end at 8~m/s, which is the range of wind values strongly represented in the data after cleaning, cf. Figure \ref{fig:power_curve}, the third column. We consider wind bins of length $0.5$. The second environmental factor that we consider is atmospheric temperature. Thus, in each wind bin, we split the observations into four smaller sub-bins corresponding to different temperature conditions.

\subsection{Study Concerning Turbine T11}
\label{subsec:detailed_empirical}

In the case study concerning turbine T11, we use at first four temperature sub-bins, with centroids at ca.~15, 18, 22, and 27 degrees Celsius. In each sub-bin, we first split the data into $R = 30$ windows, and then, we execute fuzzy c-means and extract $C = 3$ concepts. 

\begin{figure}[!h]
    \centering
    \includegraphics[width=.49\textwidth]{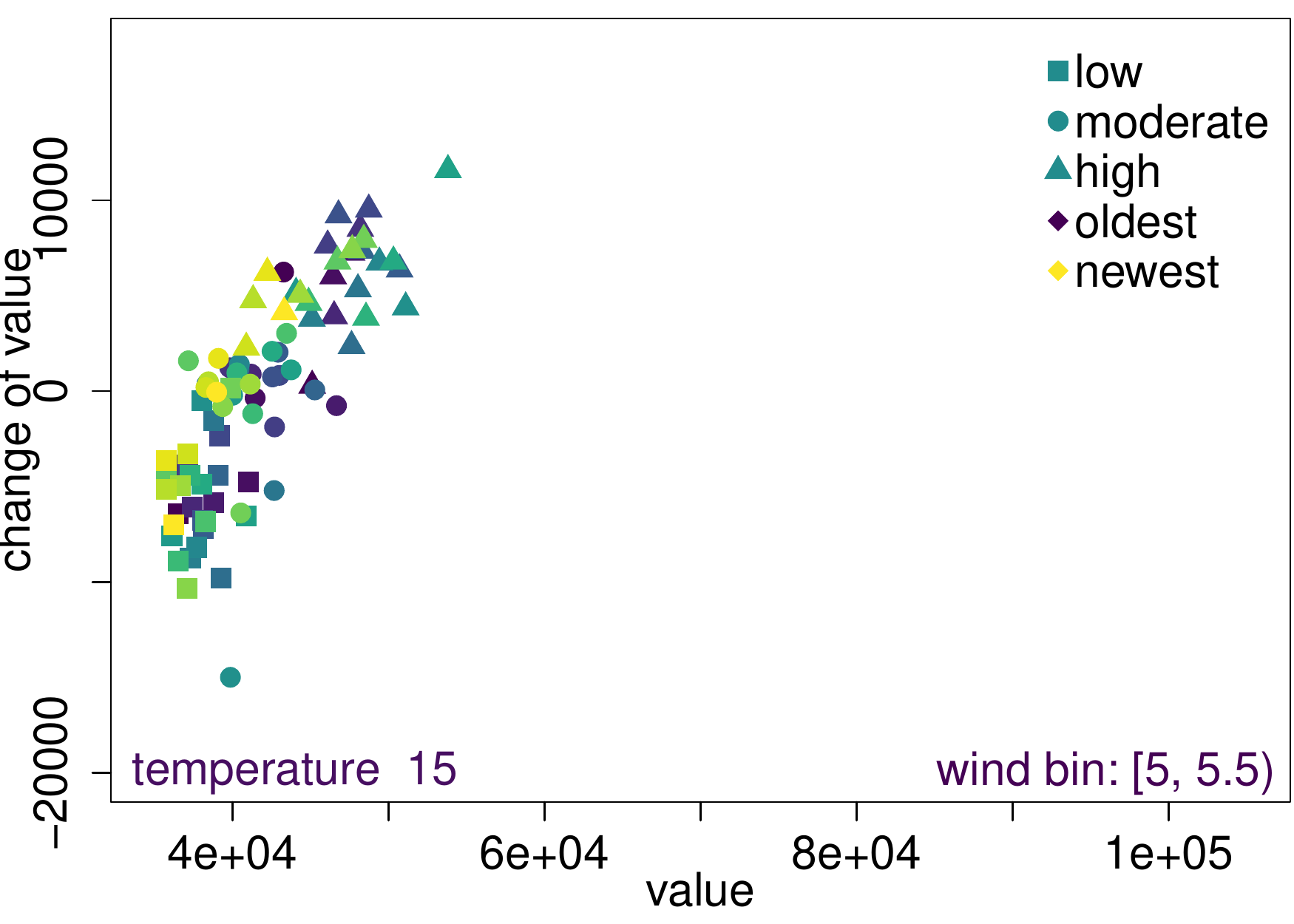}
	\includegraphics[width=.49\textwidth]{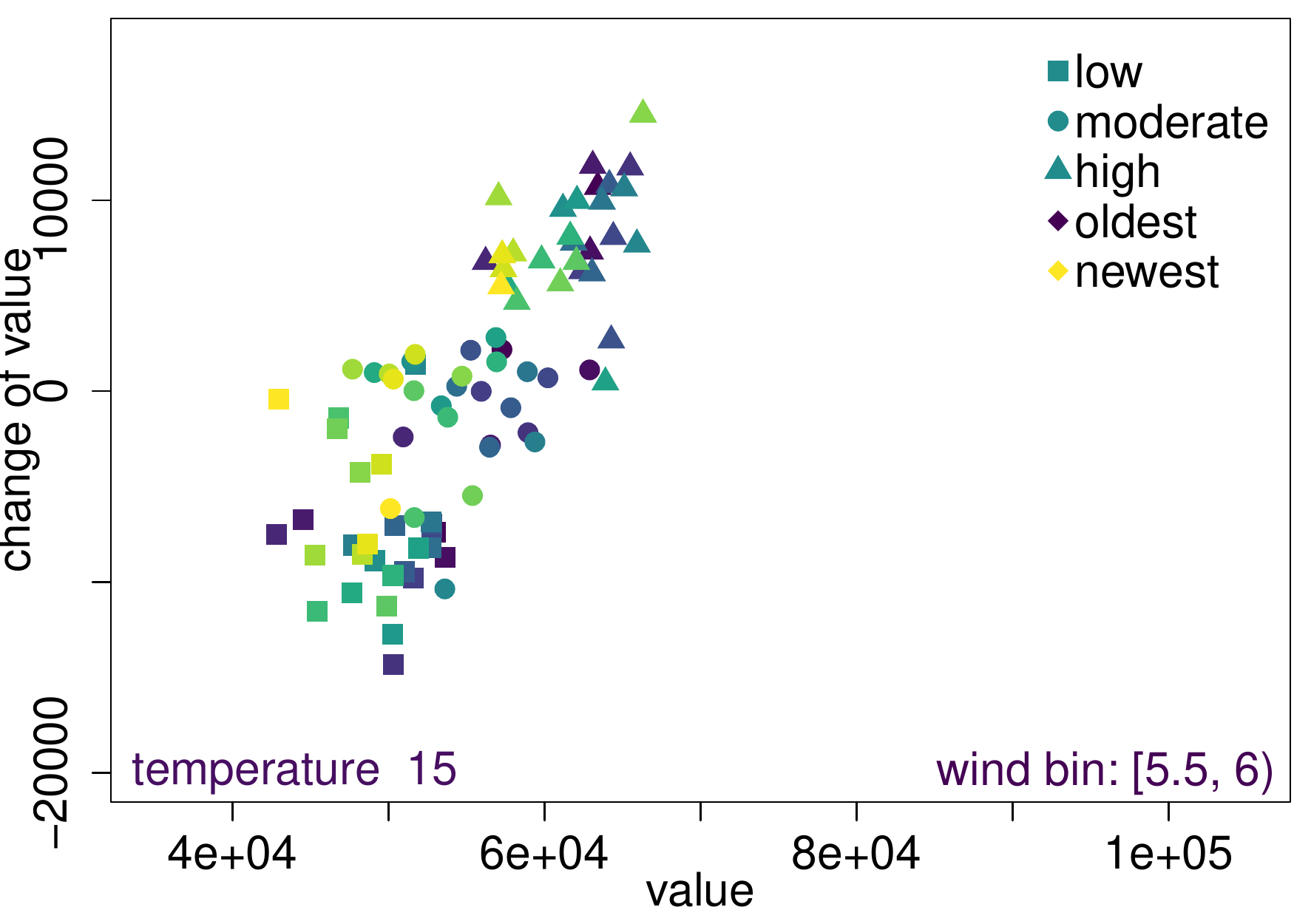}
    \includegraphics[width=.49\textwidth]{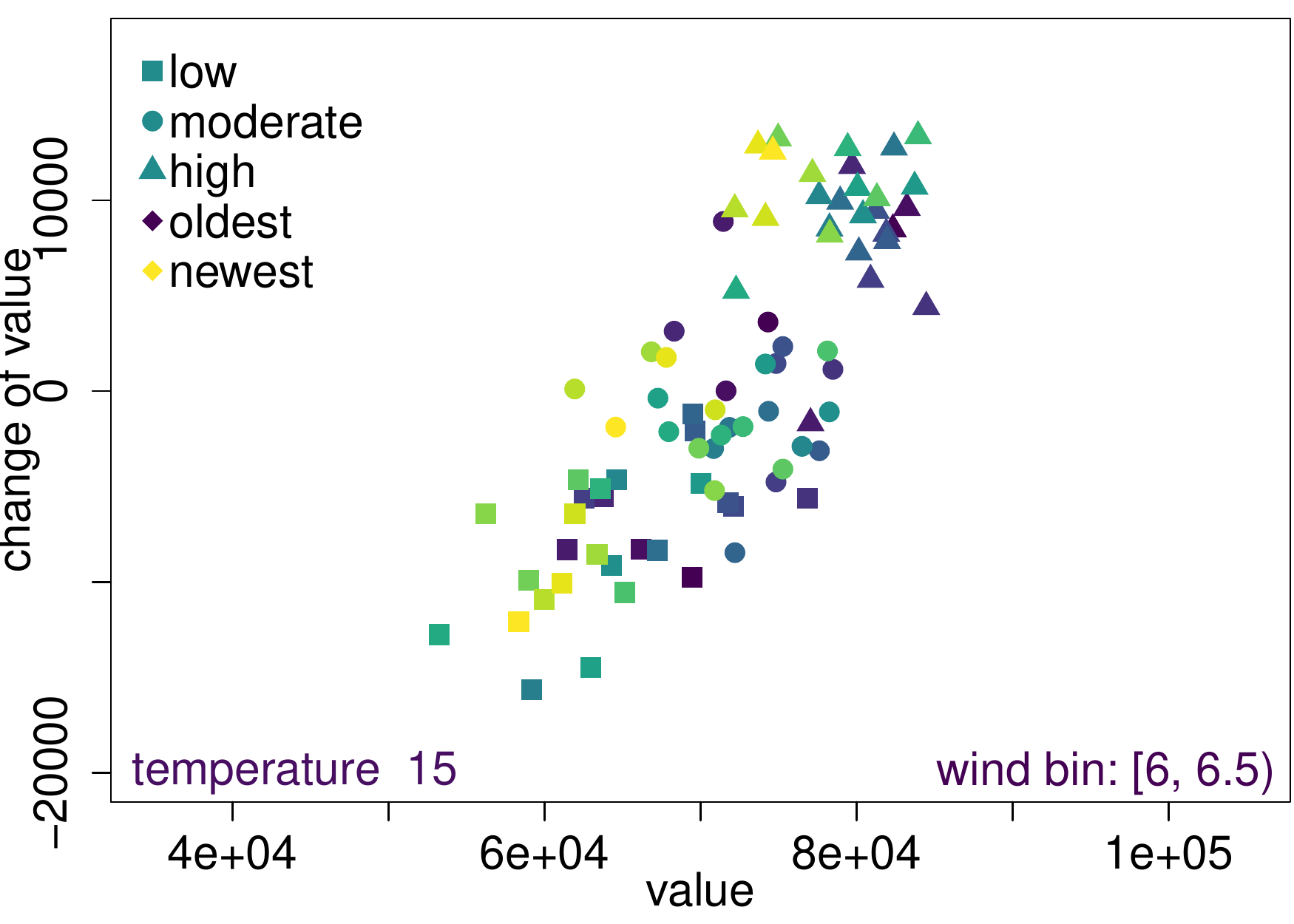}
    \includegraphics[width=.49\textwidth]{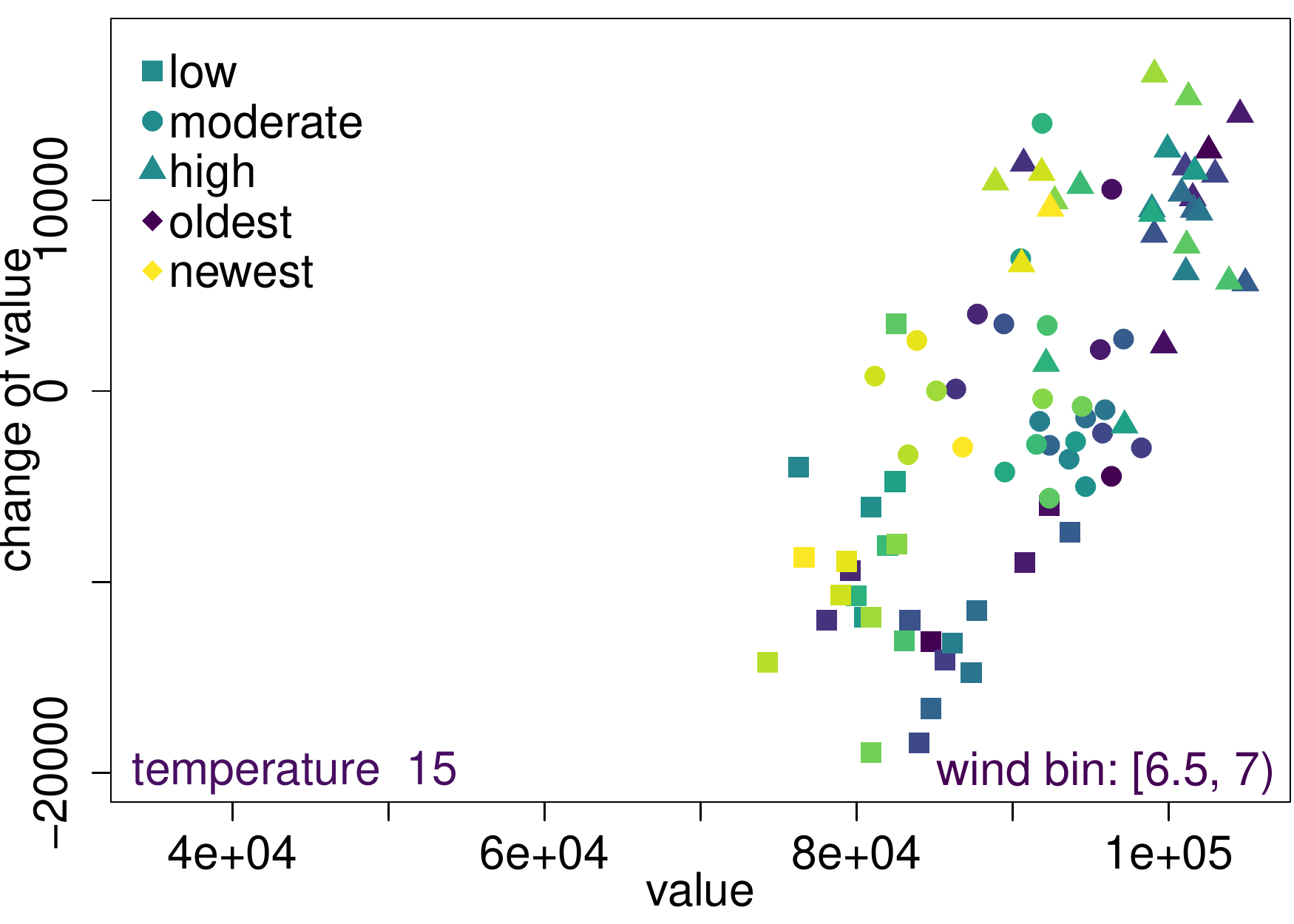}
    \caption{Concepts in the two-dimensional space of power value and change of power value turbine T11. In each window, three concepts corresponding to small (squares), moderate (circles), and high (triangles) values were extracted. Concepts are colored. The most recent are the yellow ones. In the bottom part of each plot, we placed information about wind and temperature variables. It is worth mentioning that the temperature interval is the same in each plot while the wind interval changes.}
      \label{fig:WT11_temperatures}
\end{figure}

In Figure \ref{fig:WT11_temperatures}, we display concepts extracted for different wind ranges for the turbine T11. There were three concepts extracted for each window, thus, each plot contains $3 \cdot 30$ points. Concepts were sorted according to the increasing value of produced power. Points are marked using three symbols: a~square, a~circle, and a~triangle. 
Squares depict concepts representing the smallest value, triangles correspond to the concepts that represent the highest values. Concepts introduce a~natural discretization and summarization of raw data. Concepts in Figure \ref{fig:WT11_temperatures} are colored. The yellower the color, the more recent the observation while the navy color represents the oldest observations. In the scatter plots, we observe a systematic shift of colors from the right (high power) to the left (low power) for each wind turbine. In particular, yellowish colors are on the left and blueish colors tend to be positioned on the right. This behavior is a clear indication of performance deterioration. In each cluster, values on the right correspond to higher generated power, values on the left correspond to lower power values.

Gradient coloring representing the time flow enables visual comparison of the degree of deterioration of power production. The proposed method allows illustrating the degradation of the wind turbine in a way easy to understand by domain experts. Splitting the data into sub-bins helps to analyze the differences between the machine's operating characteristics under different weather conditions. Concepts and membership to concepts can both be interpreted straightforwardly. Concepts provide a data-driven discretization of the values present in the data. Furthermore, we attach linguistic labels to describe the values represented by concepts. The use of $C=3$ concepts entails that we are describing the tendency of the power production to be low, moderate, or high. The use of human-friendly linguistic labels ``low'', ``moderate'', and ``high'' make the analyzes convenient.

Subsequently, we present the health index based on regression models. We considered wind ranges from $[5, 5.5)$ to $[7, 7.5)$ and four temperature sub-bins. In each wind and temperature sub-bin, we computed a regression model using Equation \eqref{eqn:regression_mu}. Such a regression model concerns membership values to the concepts representing high power production. Table~\ref{tab:largest_WT11} presents the obtained regression slopes such that negative values indicate a decreasing trend that entails turbine performance degradation.

\begin{table}[!ht]
\centering
\scriptsize{
\renewcommand{\arraystretch}{1.1}
\caption{Regression slope-based health index for turbine T11. The table concerns the case of concepts describing high power production. Therefore, negative values indicate aging. Models were computed separately for different wind and temperature conditions. Values were multiplied by $10^5$ to enhance clarity. }
\label{tab:largest_WT11}
\centering
\begin{tabular}{rr|r|r|r|r|r}
\hline
\hline
&&\multicolumn{5}{c}{Wind bins}	\\\cline{3-7}
&&[5, 5.5) & [5.5, 6)&[6, 6.5)&[6.5, 7)&[7, 7.5)\\
	\hline\hline
\multirow{4}{*}{\rotatebox{90}{Temp. ($^{\circ}$C)} }&	
15  &  $-$6.64  &3.71  &5.29 &$-$3.94 &$-$24.76\\
&18& $-$6.78& $-$6.69 &$-$11.79 &$-$6.99 &$-$35.98\\
&22  &$-$1.34 &$-$18.67 &$-$2.99 &$-$0.31  &5.75\\
&27  &0.61& 5.62 &$-$12.03 & 10.09 &$-$1.13\\\hline
&\textbf{sum}&\textbf{$-$14.15	}&\textbf{$-$16.03}&	\textbf{$-$21.52	}&\textbf{$-$1.15	}&\textbf{$-$56.12} \\
\hline
\hline
\end{tabular}
}
\end{table}

\begin{figure}[!htbp]
	\centering 
	\includegraphics[width=0.6\textwidth]{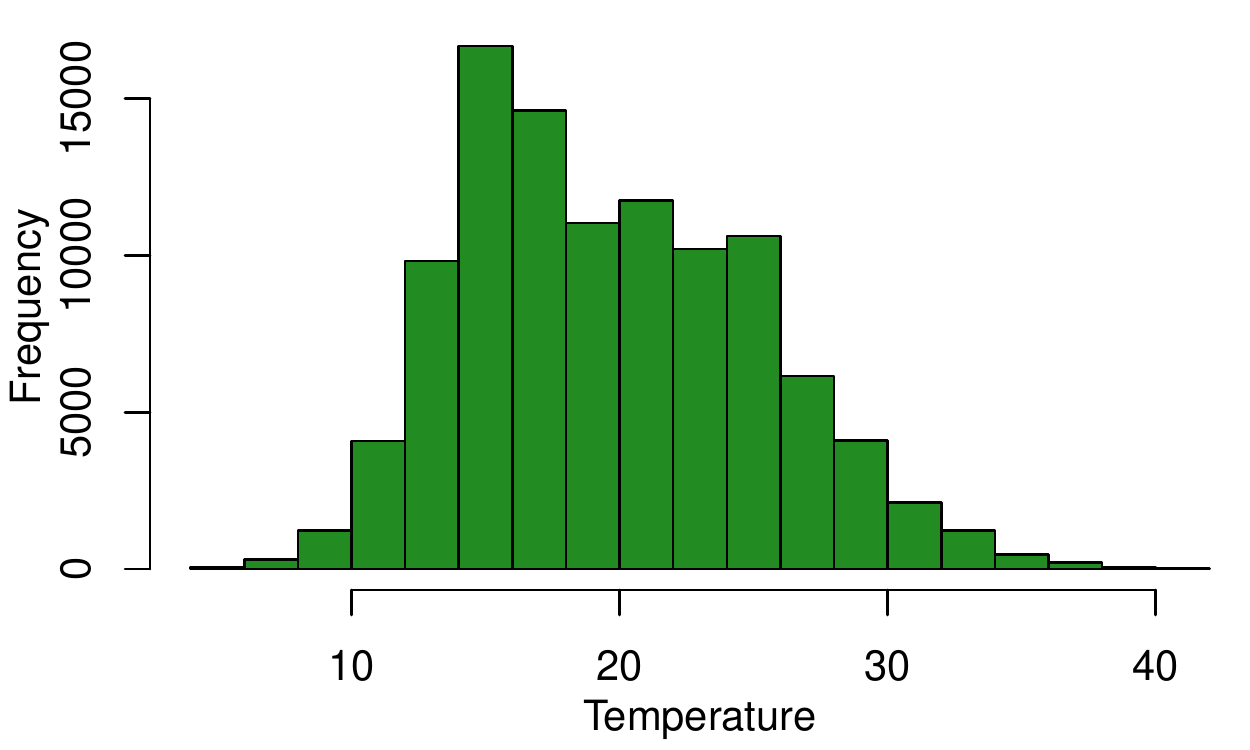} 
	\caption{Histogram of temperature values collected at the wind turbine.}
	\label{fig:temperature}
\end{figure}

In Table~\ref{tab:largest_WT11}, we observe negative slopes in most cases. It is challenging to interpret the results concerning the highest temperatures, around 27$^\circ$C. In this case, slopes are more often positive, when looking at the temperature values histogram in Figure \ref{fig:temperature}, we notice that these values are close to extreme operating conditions for this machine.  

Table \ref{tab:smallest_WT11} presents the regression slopes of models computed for membership values to the low power production concepts. They were computed according to Equation \eqref{eqn:regression_mu3}. In this case, positive values indicate that there is an undesired shift. In other words, positive values show that T11 started to output lower power values with time. In Table~\ref{tab:smallest_WT11}, in the vast majority of cases (i.e., in the majority of analyzed wind and temperature sub-bins), the result indicates aging, because values have a positive sign. 

\begin{table}
\centering
\scriptsize{
\renewcommand{\arraystretch}{1.1}
\caption{Regression-based health index: slopes of regression models fitted to memberships to concepts that represent low power production. Positive signs indicate aging. Results concern selected wind and temperature conditions. Slopes were multiplied by $10^5$. }
\label{tab:smallest_WT11}
\centering
\begin{tabular}{rr|r|r|r|r|r}
\hline
\hline
&&\multicolumn{5}{c}{Wind bins}\\\cline{3-7}
&&[5, 5.5) & [5.5, 6)&[6, 6.5)&[6.5, 7)&[7, 7.5)\\
\hline\hline
\multirow{4}{*}{\rotatebox{90}{Temp. ($^{\circ}$C)} }&	
15	&	$-$5.29	&	$-$1.47	&	$-$3.44	&	4.80	&	28.54\\
&18	&	$-$4.73	&	6.96	&	11.11	&	$-$4.26	&	25.03\\
&22&	0.79	&	7.94	&	$-$0.53	&	$-$7.77	&	$-$1.34\\
&27	&	31.34	&	2.91	&	0.47	&	$-$7.86	&	$-$11.97\\\hline
&\textbf{sum}&\textbf{22.12}	&	\textbf{16.34}	&	\textbf{7.62}	&	\textbf{$-$15.09}	&\textbf{40.26}	\\
\hline
\hline
\end{tabular}
}
\end{table}

\subsection{Comparative Analysis}
\label{subsec:comparative_empirical}
 
Subsequently, we compare regression model-based health indexes computed for all wind turbines in the EDP data set. In this experiment, we still focus on four temperature sub-bins around 15, 18, 22, and 27 degrees Celsius. We do not recommend increasing the number of temperature sub-bins since it reduces the number of data points in each sub-bin (the sub-bins get smaller). In these experiments, we set $C = 3$ and we change the parameter $R$ to 10. Health indexes for each turbine are given in Tables~\ref{tab:largest_all_temp} and~\ref{tab:smallest_all_temp} together with row-wise sums. They concern membership to the high and the low power production concepts, respectively.

Table~\ref{tab:largest_all_temp} presents the slopes of the regression function fitted to membership values concerning concepts describing high power production values. The table contains results for different weather conditions for all wind turbines in the EDP data set. In many cases, we obtained a negative slope indicating decreasing health of the studied objects. Numerical values are comparable for each turbine since they were computed with the same settings. Therefore, we may infer that T01 deteriorated the least, while T07 deteriorated the most. The faster the wind, the more visible is the deterioration.

\begin{table}[!ht]
\centering
\scriptsize{
\renewcommand{\arraystretch}{1.1}
\caption{Regression model coefficients computed for membership values to the concepts describing high power production and $R = 10$. We studied four temperature sub-bins and five wind ranges. Negative signs indicate aging. Values were multiplied by $10^5$. }
\label{tab:largest_all_temp}
\centering
\begin{tabular}{c|c|r|r|r|r|r}
\hline
\hline
\multicolumn{1}{c|}{Turbine}&Temp.&\multicolumn{5}{c}{Wind bins (m/s)} 	\\\cline{3-7}
\multicolumn{1}{c|}{ID}&\multicolumn{1}{c|}{($^\circ$C)}&[5, 5.5) & [5.5, 6)&[6, 6.5)&[6.5, 7)&[7, 7.5)
\\
	\hline\hline
\multirow{5}{*}{T01}&	15 &	7.49	&3.46&	$-$0.70	&1.65&	30.14	
\\
&18	&5.02	&$-$9.39	&$-$23.84	&$-$3.46&	$-$24.70	
\\
&22	&$-$11.13	&4.86&	$-$11.45	&6.60	&$-$23.09	
\\
&27&	31.82	&35.94	&22.93&	6.16	&$-$69.74	
\\\cline{2-7}
&\textbf{sum}	&\textbf{33.20}	&\textbf{34.87}&	\textbf{$-$13.06}	&\textbf{10.95}	&\textbf{$-$87.39}
\\
\hline\hline
\multirow{5}{*}{T06}&	15 &$-$5.16	&$-$1.92	&5.33&	$-$2.38&	$-$207.84	
\\
&18	&$-$16.24	&12.03	&9.30	&5.55	&$-$93.16
\\
&22	&18.62	&$-$15.73	&2.52	&19.74	&$-$0.97	
\\
&27&$-$18.13	&12.08&	$-$3.98&	54.72&	$-$138.53
\\
\cline{2-7}
&\textbf{sum}	&\textbf{-20.91}&\textbf{6.46}&	\textbf{13.17}&	\textbf{77.63}	&\textbf{$-$440.50}
\\
\hline
\hline
\multirow{5}{*}{T07}&	15 &$-$9.75&	8.17&	$-$6.15&	$-$7.58	&$-$185.33
\\
&18	&$-$1.62&	$-$12.84	&$-$0.42&	2.20&	$-$147.45
\\
&22	&$-$7.46	&$-$4.50	&$-$14.56	&$-$12.69&	$-$114.85
\\
&27& 4.33	&$-$26.02&	36.28&	24.59&	$-$78.23
\\\cline{2-7}
&\textbf{sum}	&\textbf{$-$14.50}&\textbf{$-$35.19}&\textbf{15.15}&\textbf{6.52}&\textbf{$-$525.86}
\\
\hline
\hline
\multirow{5}{*}{T11}&	15 &0.10	&9.71&	1.07&	$-$5.27&	$-$42.69
\\
&18	&$-$9.17&	$-$15.26&	5.81&	14.72&	$-$50.96
\\
&22	&$-$4.64	&$-$23.66&	$-$1.52&	$-$16.08&	$-$2.17
\\
&27&10.96	&$-$10.08	&$-$20.96&	$-$83.84&	$-$3.51
\\\cline{2-7}
&\textbf{sum}	&\textbf{$-$2.75}	&\textbf{$-$39.29	}&\textbf{$-$15.60}	&\textbf{$-$90.47}	&\textbf{$-$99.33}
\\
\hline
\hline
\end{tabular}
}
\end{table}

In Table~\ref{tab:smallest_all_temp}, we present health indexes computed for all wind turbines in the EDP data set concerning low power production. Indexes tend to have a positive sign which confirms performance degradation. The largest slopes concern the $[7,7.5)$ wind range.

\begin{table}[!ht]
\centering
\scriptsize{
\renewcommand{\arraystretch}{1.1}
\caption{Regression model coefficients computed for membership values to the concepts describing low power production and $R = 10$.  We use four temperature sub-bins and five wind ranges. Positive coefficients indicate aging. Values were multiplied by $10^5$. }
\label{tab:smallest_all_temp} 
\centering
\begin{tabular}{c|c|r|r|r|r|r}
\hline
\hline
\multicolumn{1}{c|}{Turbine}& Temp.&\multicolumn{5}{c}{Wind bins (m/s)} 	\\\cline{3-7}
\multicolumn{1}{c|}{ID}&\multicolumn{1}{c|}{($^\circ$C)}&[5, 5.5) & [5.5, 6)&[6, 6.5)&[6.5, 7)&[7, 7.5)\\
	\hline\hline
\multirow{5}{*}{T01}&	15 &$-$4.44	&9.92	&2.28	&$-$5.75	&$-$6.62	
\\
&18	&$-$4.15&	18.65&	15.46	&$-$1.49&	0.24
\\
&22	&6.58	&$-$9.67	&7.56	&$-$6.51	&3.42
\\
&27&46.01&	2.18	&$-$33.05	&20.79	&25.07
\\\cline{2-7}
&\textbf{sum}	&\textbf{44.00}&	\textbf{21.08}	&\textbf{$-$7.75}	&\textbf{7.04}&	\textbf{22.11}	
\\
\hline\hline
\multirow{5}{*}{T06}&	15 &$-$0.64	&8.60	&$-$4.45	&$-$10.88&	91.83
\\
&18	&4.70	&$-$11.88	&$-$0.22	&$-$11.41	&65.37
\\
&22	&$-$11.44&	26.30&	$-$4.07&	$-$14.05	&60.53
\\
&27&0.51	&$-$19.64&	$-$3.60&	$-$40.47	&48.70
\\\cline{2-7}
&\textbf{sum}	&\textbf{$-$6.87}	&\textbf{3.38}&	\textbf{$-$12.34}	&\textbf{$-$76.81}&	\textbf{266.43}
\\
\hline
\hline
\multirow{5}{*}{T07}&15&6.41&	$-$4.31&	10.45	&$-$3.39	&78.83	
\\
&18	&0.83	&19.26	&4.39	&16.17&	98.68
\\
&22	&3.00	&1.47&	21.98&	8.77	&61.85	
\\
&27&$-$25.91	&23.40	&$-$14.86&	$-$13.95&	$-$176.66	
\\ \cline{2-7}
&\textbf{sum}	&\textbf{$-$15.67}&	\textbf{39.82}&	\textbf{21.96}	&\textbf{7.60}&\textbf{62.70}
\\
\hline
\hline
\multirow{5}{*}{T11}&	15 &10.67	&$-$1.83	&$-$8.12&	7.52&	57.14
\\
&18	&$-$15.90	&7.86	&3.33&	$-$28.14&	56.68	
\\
&22	&8.10&0.42&	$-$0.02&	$-$4.44&	7.21
\\
&27&10.04	&$-$5.51	&1.14&	53.18	&$-$24.97
\\\cline{2-7}
&\textbf{sum}	&\textbf{12.91}&	\textbf{0.94}&	\textbf{$-$3.67}&	\textbf{28.12}	&\textbf{96.06}	
\\
\hline
\hline
\end{tabular}
}
\end{table}

Similarly to the previous experimental setup, aging is visible for $R = 10$. In Table~\ref{tab:largest_all_temp}, we can observe that memberships to the concepts describing high power production are declining with time. This means that high power production declines with time. In contrast, as it is illustrated in Table~\ref{tab:smallest_all_temp}, memberships to the concepts describing small power production are increasing with time (slopes have positive signs). In other words, for fixed environmental conditions, we observe the degradation of the power production process. In Tables~\ref{tab:largest_all_temp} and~\ref{tab:smallest_all_temp}, we can see that the T01 turbine is the least affected by the aging processes, while the T06 and T07 turbines are the most affected.

\begin{figure}
     \centering
     
     \begin{subfigure}{0.22\textwidth}
         \centering
         \includegraphics[page=1,scale=0.4,trim={0 0 0 0}, clip]{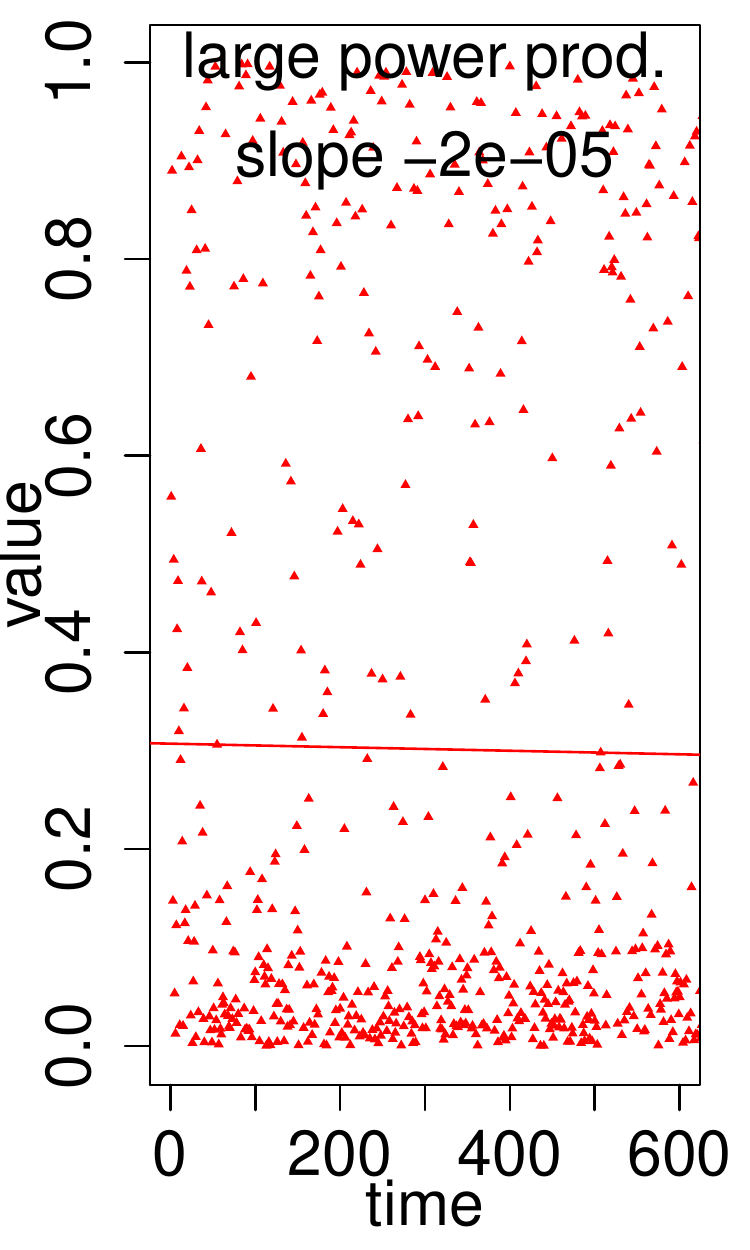}
         \caption{T01}
     \end{subfigure}
     \hfill
     \begin{subfigure}{0.22\textwidth}
         \centering
         \includegraphics[page=1,scale=0.4,trim={0 0 0 0}, clip]{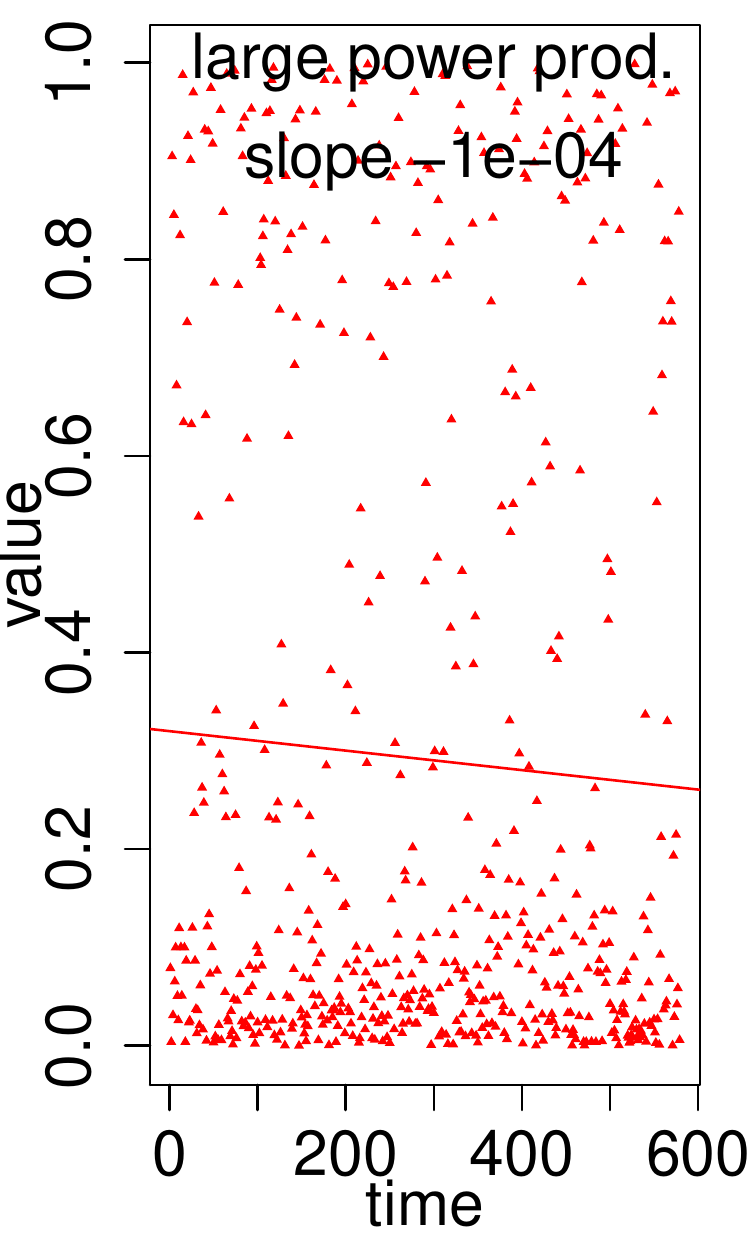}
         \caption{T06}
     \end{subfigure}
     \hfill
     \begin{subfigure}{0.22\textwidth}
         \centering
         \includegraphics[page=1,scale=0.4,trim={0 0 0 0}, clip]{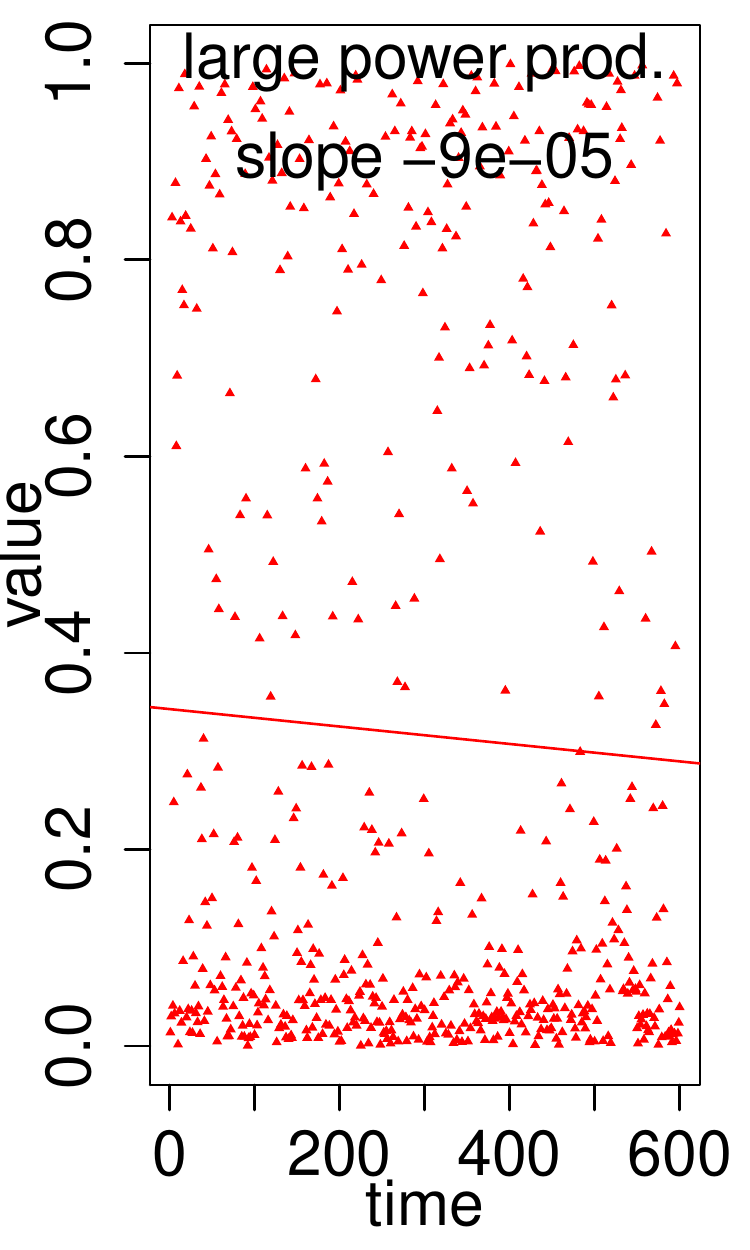}
         \caption{T07}
     \end{subfigure}
     \hfill
      \begin{subfigure}{0.22\textwidth}
          \centering
          \includegraphics[page=1,scale=0.4,trim={0 0 0 0}, clip]{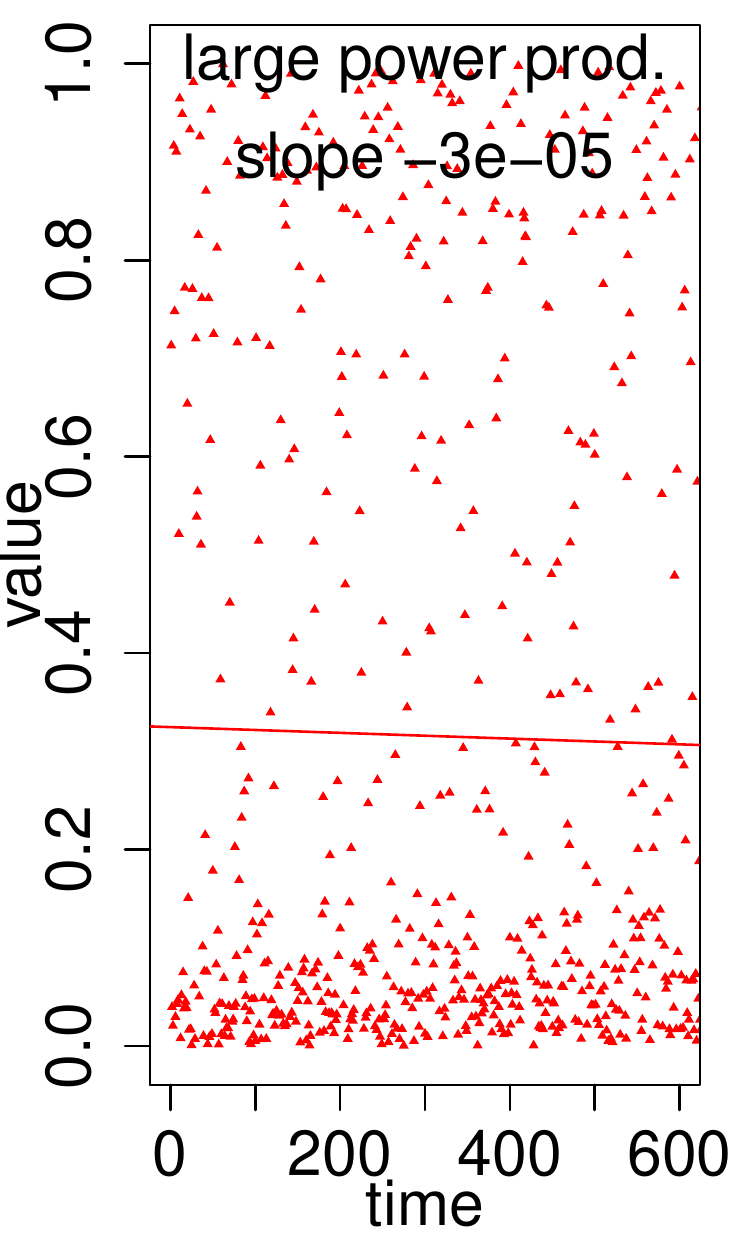}
          \caption{T11}
      \end{subfigure} 
      \hfill
	 \begin{subfigure}{0.22\textwidth}
         \centering
         \includegraphics[page=1,scale=0.4,trim={0 0 0 0}, clip]{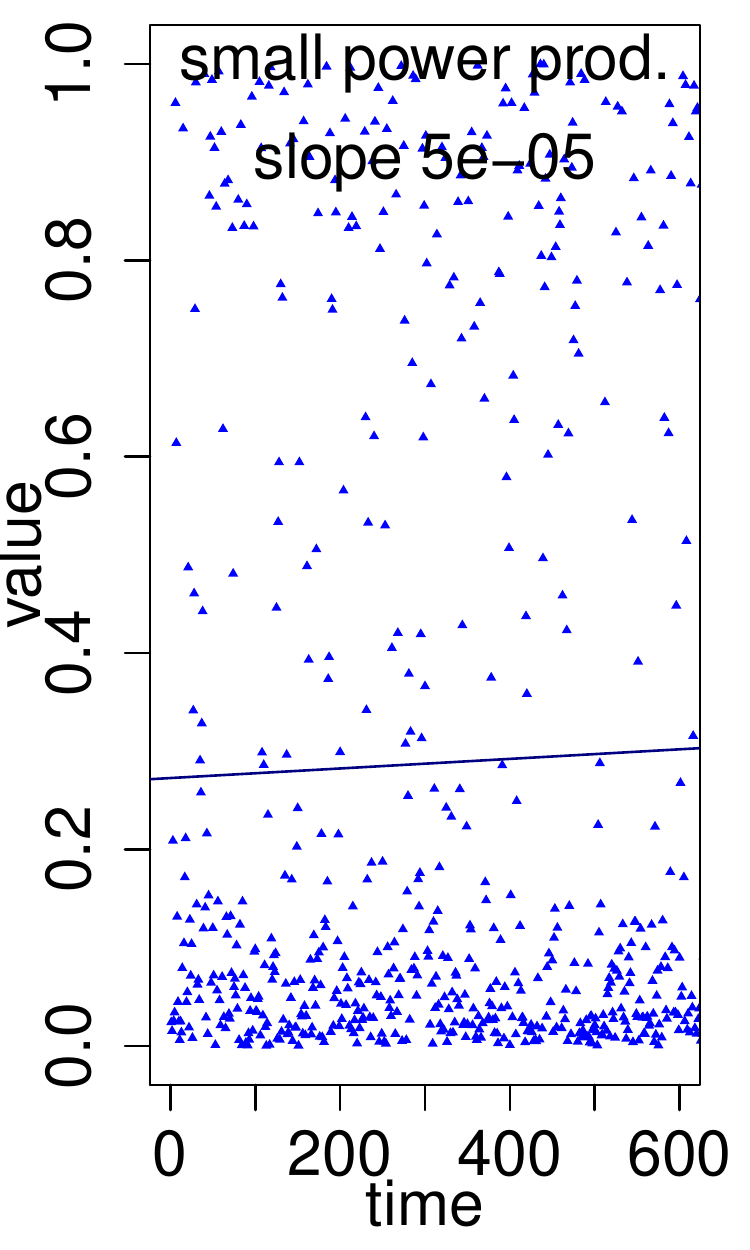}
         \caption{T01}
     \end{subfigure}
     \hfill
     \begin{subfigure}{0.22\textwidth}
         \centering
         \includegraphics[page=1,scale=0.4,trim={0 0 0 0}, clip]{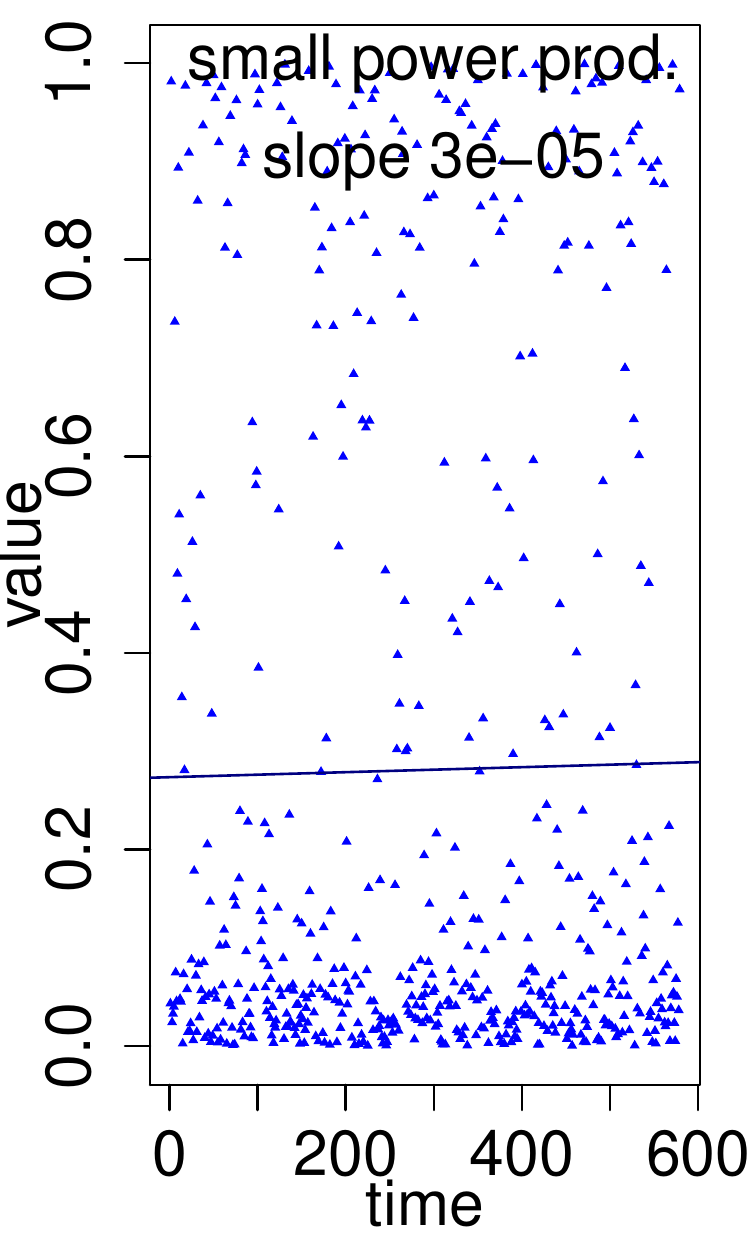}
         \caption{T06}
     \end{subfigure}
     \hfill
     \begin{subfigure}{0.22\textwidth}
         \centering
         \includegraphics[page=1,scale=0.4,trim={0 0 0 0}, clip]{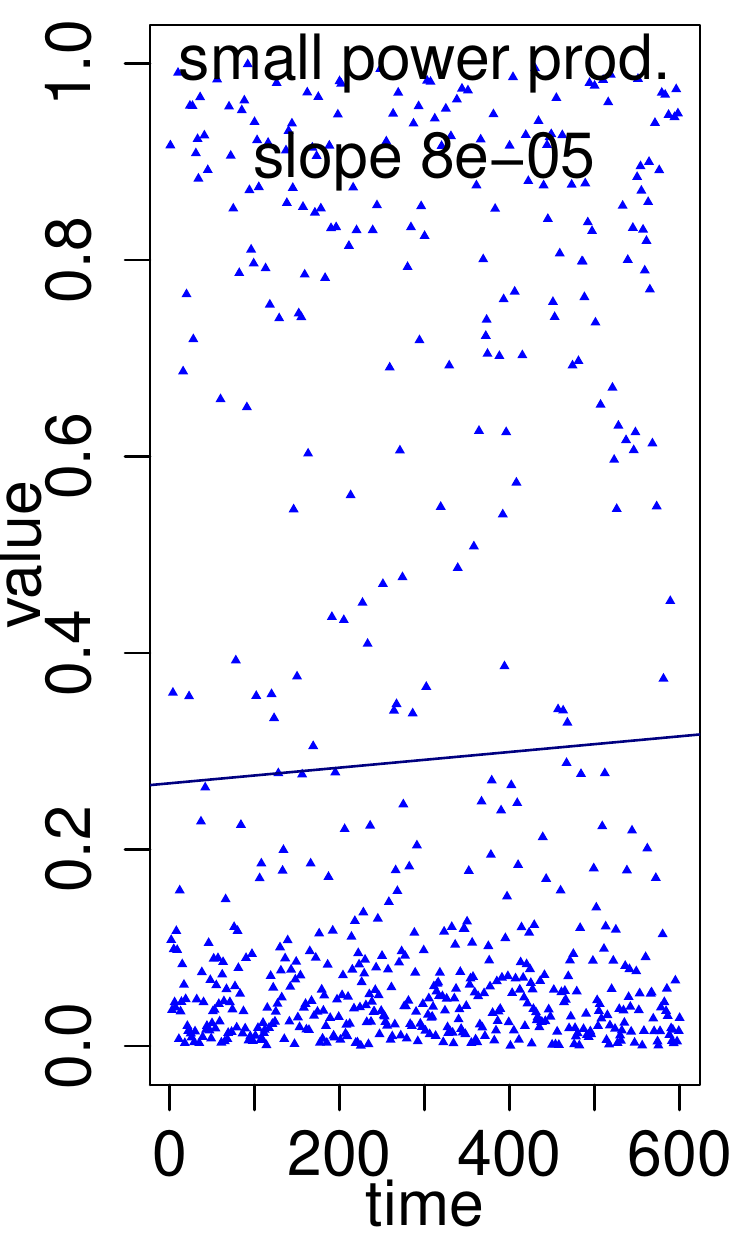}
         \caption{T07}
     \end{subfigure} 
     \hfill
      \begin{subfigure}{0.22\textwidth}
          \centering
          \includegraphics[page=1,scale=0.4,trim={0 0 0 0}, clip]{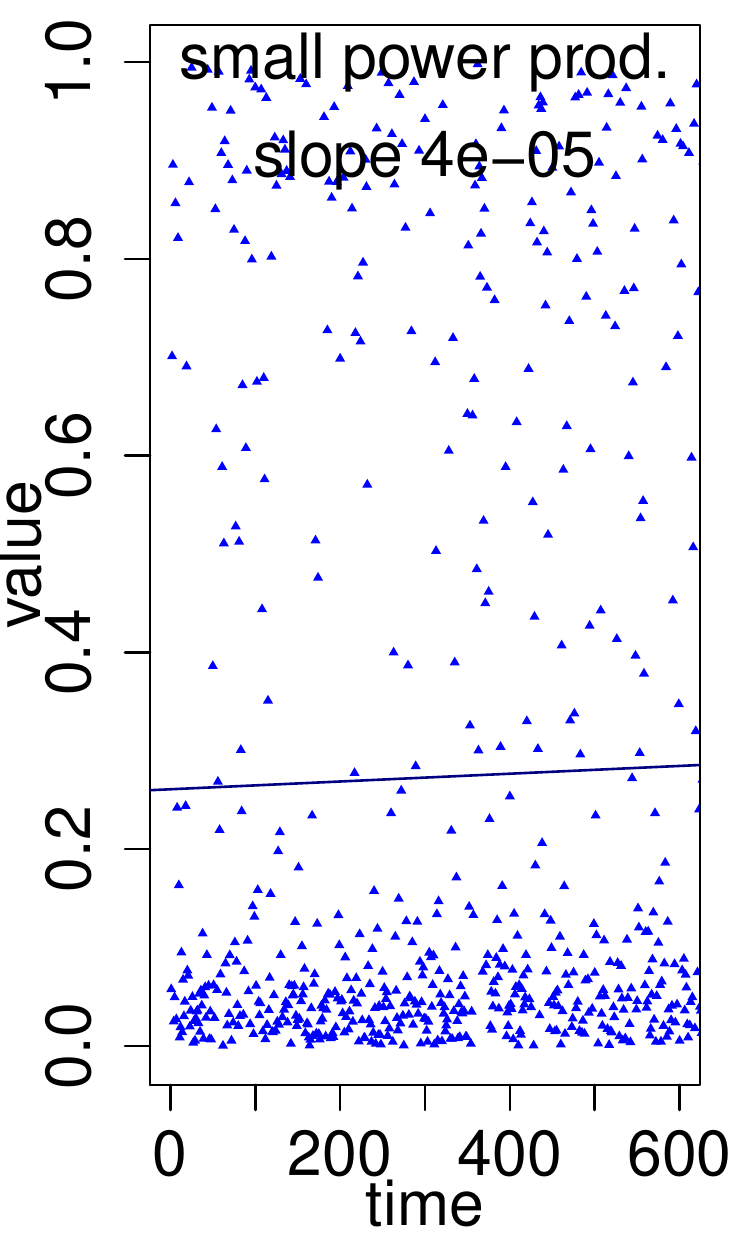}
          \caption{T11}
      \end{subfigure}      
   \hfill      
      \caption{Example slopes computed for four wind turbines for the same weather conditions. Top row concerns memberships to the concepts of high power production. Here, negative values indicate aging. Bottom row concerns memberships to the concepts of low power production, where positive sign of slope shows aging. The experiment concerns wind in the range $[6.5,7)$ and temperature around 15 deg. Celsius. $R = 30$. } 
      \label{fig:slopes_on_mmeberships}
\end{figure}

Figure \ref{fig:slopes_on_mmeberships} presents plots concerning membership values ordered in time, as a way to visualize the results. The top row concerns memberships to the large power production concepts. The bottom row concerns memberships to the low power production concepts. In both cases, we have fitted linear regression models and the obtained lines are visible in the plots. We adopt the slopes of the obtained linear models to inspect turbine performance degradation. Negative signs of regression models in the first row indicate aging (they tell that high power production declines with time). In contrast, positive signs in the bottom row indicate aging: low power production increases with time. The experiment concerned wind in the $[6.5,7)$ range  and temperature around 15 degrees Celsius for all wind turbines.

\begin{figure}
     \centering
     \begin{subfigure}[b]{0.48\textwidth}
         \centering
         \includegraphics[width=\textwidth]{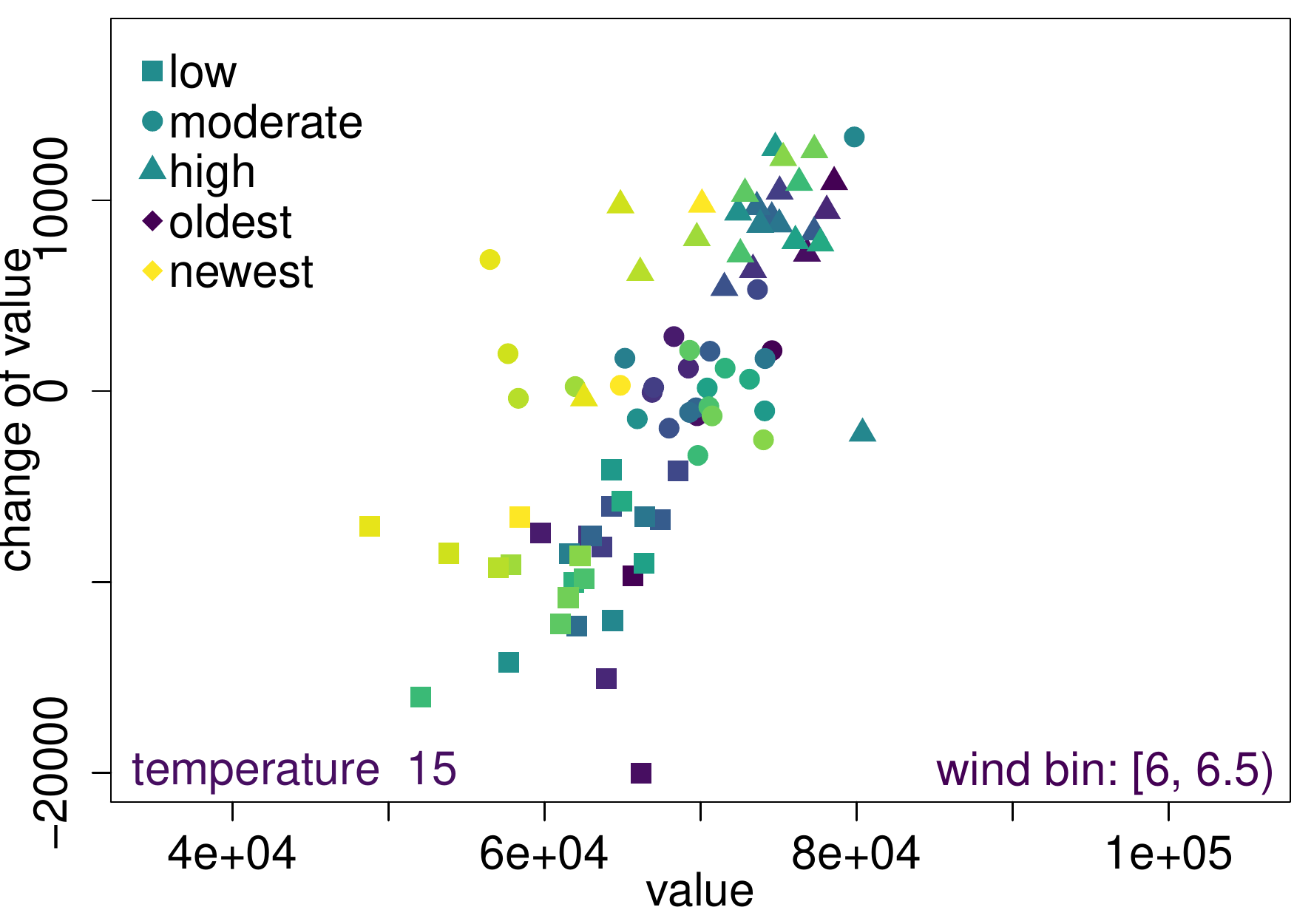} 
         \caption{T01}
         \label{fig:comparative_towersT01}
     \end{subfigure}
     \hfill
     \begin{subfigure}[b]{0.48\textwidth}
         \centering
         \includegraphics[width=\textwidth]{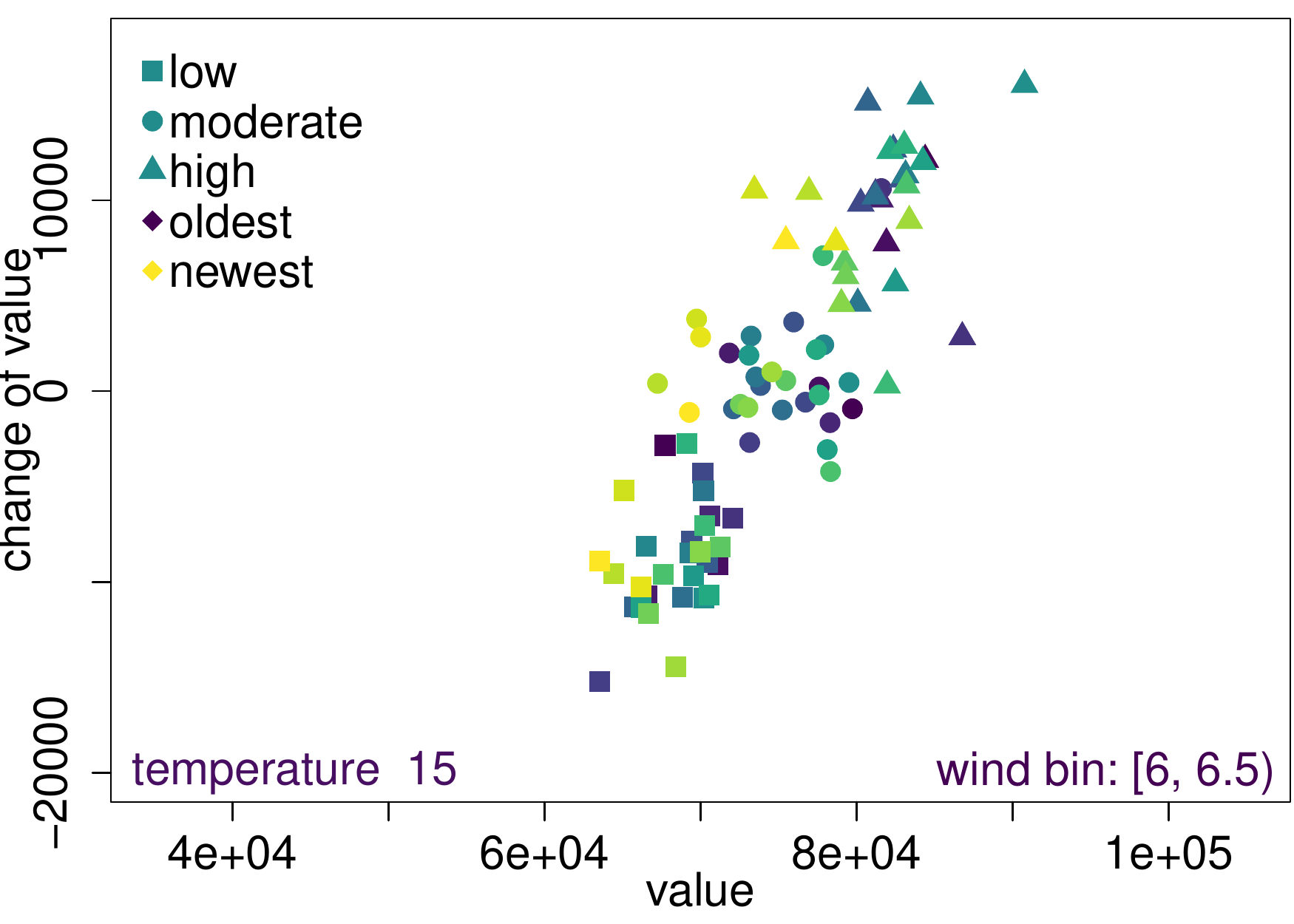}
         \caption{T06}
         \label{fig:comparative_towersT06}
     \end{subfigure}
     \hfill
     \begin{subfigure}[b]{0.48\textwidth}
         \centering
         \includegraphics[width=\textwidth]{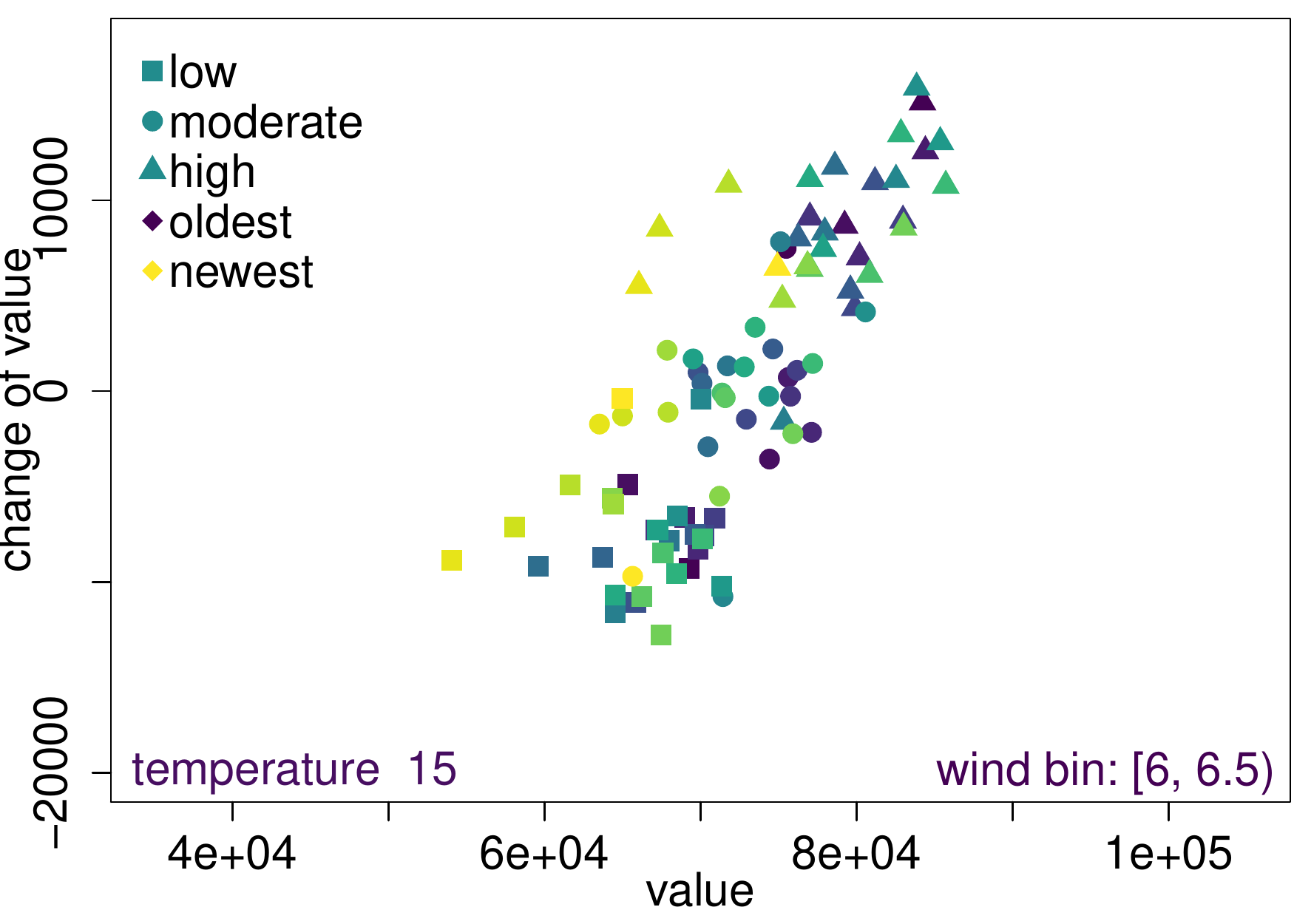}
         \caption{T07}
         \label{fig:comparative_towersT07}
     \end{subfigure}
      \caption{Concepts in the two-dimensional space of power value and change of power value. For each window, three concepts were extracted corresponding to low (squares), moderate (circles), and high (triangles) values. Concepts are colored. The most recent are the yellow ones. Tower IDs are given in individual captions. Plots concern wind values between 6 and 6.5 m/s, temperature around 15 deg.~C. The plot for T11 is in Figure \ref{fig:WT11_temperatures}. }
      \label{fig:comparative_towers}
\end{figure}

Figure \ref{fig:comparative_towers} presents concept-based models created for turbines T01, T06, and T07. For the sake of consistency in the simulations, we used the same experimental settings used to produce plots in Figure \ref{fig:WT11_temperatures}. Notice that we do not repeat the plot for T11 to avoid redundancy. 

\subsection{Additional Visual Aid to Evaluate Aging}
\label{subsec:addition}
 
Finally, we would like to address the second method of turbine aging evaluation related to the visualization of the concepts in the \mbox{two-dimensional} space of value and change of value. 
Let us recall that for each window (we have $R$ windows), we created three clusters to represent the underlying raw values.  The second proposed method uses extracted centroids and runs clustering for these centroids. Fuzzy c-means is executed separately three times: for centroids representing low, moderate, and high power production and each time two new centroids are produced. The new centroids can be used to describe a discrepancy in the low, moderate, and high power production. For each pair of new centroids, the further apart they are, the more differentiation we observed with time. We measured this differentiation with the Distance Index (DI) given in Equation \eqref{eqn:dist_index}.

\begin{figure}[!htbp]
  \centering
     \begin{subfigure}[b]{0.49\textwidth}
         \centering
         \includegraphics[width=\textwidth]{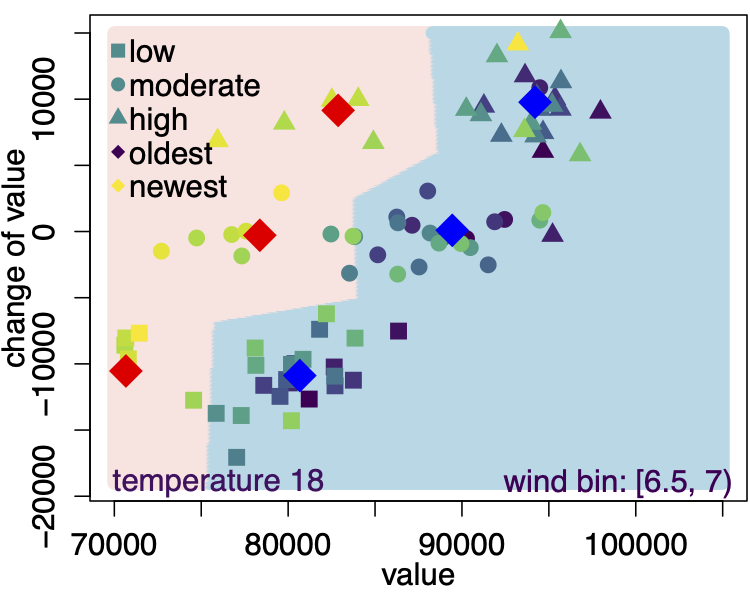} 
         \caption{T01, $DI = 0.59$}
         \label{fig:comparative_visualsT01}
     \end{subfigure}
     \hfill
     \begin{subfigure}[b]{0.49\textwidth}
         \centering
         \includegraphics[width=\textwidth]{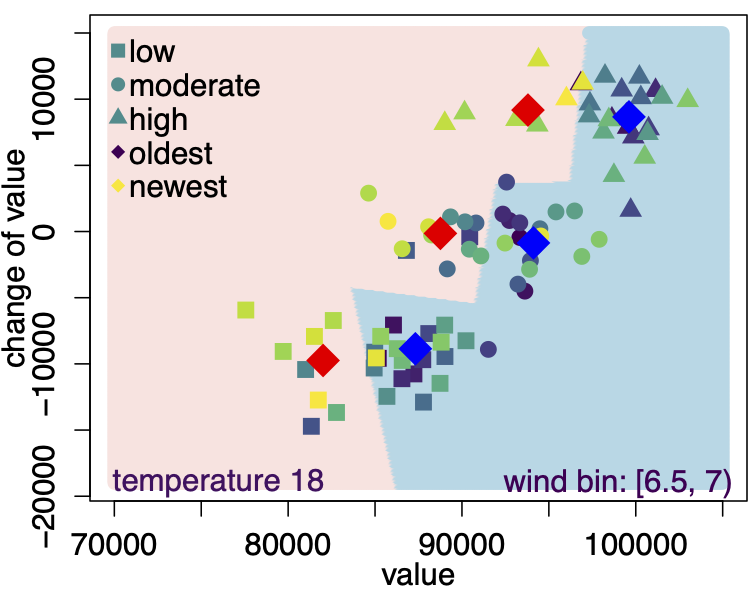}
         \caption{T06, $DI = 0.29$}
         \label{fig:comparative_visualsT06}
     \end{subfigure}
     \hfill
     \begin{subfigure}[b]{0.49\textwidth}
         \centering
         \includegraphics[width=\textwidth]{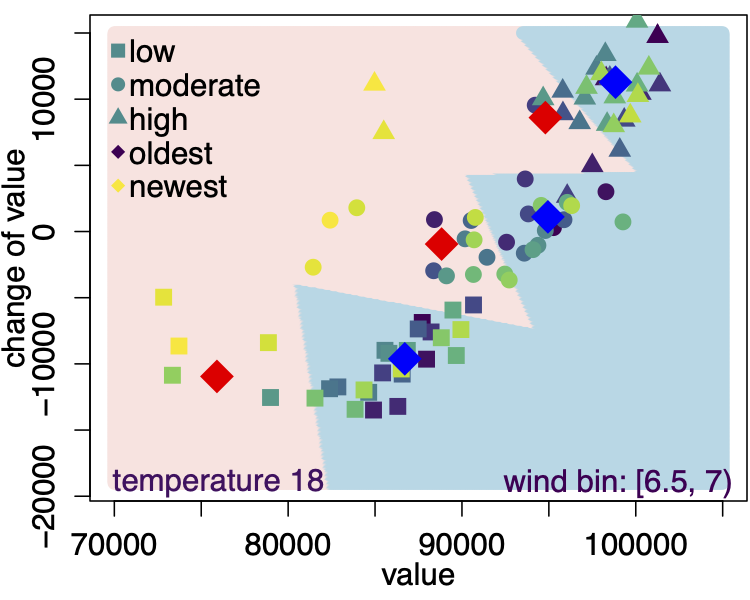}
         \caption{T07,  $DI = 0.39$}
         \label{fig:comparative_visualsT07}
     \end{subfigure}
      \hfill
          \begin{subfigure}[b]{0.49\textwidth}
              \centering
              \includegraphics[width=\textwidth]{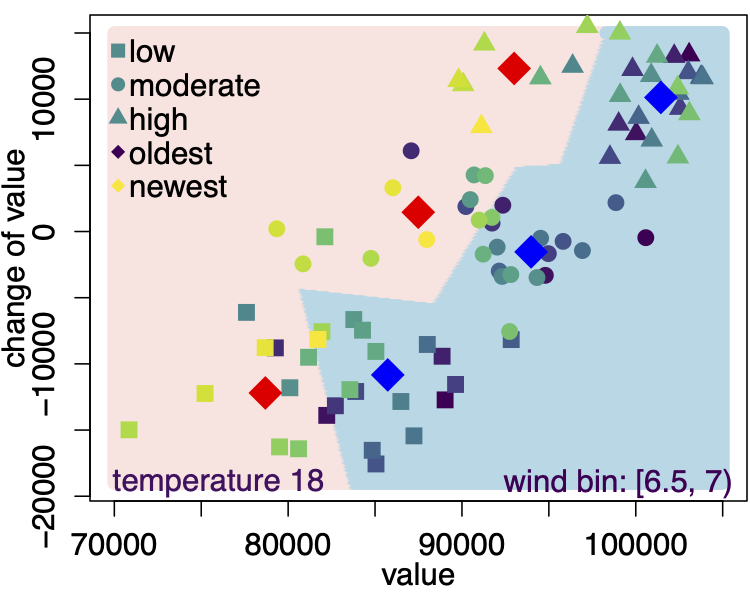}
              \caption{T11, $DI = 0.40$}
              \label{fig:comparative_visualsT11}
          \end{subfigure}
      \caption{Secondary clustering applied to concepts extracted for four wind turbines. Turbine ID is given in sub-captions. The experiment concerned $R = 20$, $C = 3$, four temperature sub-bins. Distance Index given by Equation \eqref{eqn:dist_index} and normalized according to Equation \eqref{eqn:normalization} are displayed in sub-captions.}
      \label{fig:comparative_visuals}
\end{figure}

In Figure \ref{fig:comparative_visuals}, we present plots concerning the same wind speed and temperature conditions. We examine the performance of all four wind turbines in the EDP data set. Concepts used to compute the DI (Equation \eqref{eqn:dist_index}) are marked with red and blue color diamonds. Blue is used to denote concepts describing high power production ($\textbf{v}_H^{(1)}, \textbf{v}_H^{(2)}, \textbf{v}_H^{(3)}$) and red is used to denote concepts describing low power production ($\textbf{v}_L^{(1)}, \textbf{v}_L^{(2)}, \textbf{v}_L^{(3)}$). We colored the plot background so that for a given point in the plotted coordinate system the color informs which kind of concept is the closest. Pink indicates that the closest concept is red (low power), while light blue informs that the closest concept is blue (high power production). In addition, we report the DI values computed according to Equation \eqref{eqn:dist_index} in the caption of each figure. These values were normalized according to the formula:\vspace{-5pt}
\begin{equation}
    x_i=\frac{x_i - \min}{\max-\min},
\label{eqn:normalization}
\end{equation}

\noindent where as $\min$ we used $40,000$ for the power value and $-20,000$ for the change of power for all wind turbines, both measured in Watt-hour. The change of value can be negative. As $\max$ we used $100,000$ for the power value and $20,000$ for the change of power value for all wind turbines.

It is worth recalling that the DI metric is more informative when it is computed for more than one wind-temperature sub-bin. Table~\ref{tab:distances_all} shows the normalized DI values computed for a range of wind speed and temperature sub-bins for all wind turbines in the data set.

\begin{table}[!ht]
\centering
\scriptsize{
\renewcommand{\arraystretch}{1.1}
\caption{Normalized DI values computed for four wind turbines for different wind and air temperature sub-bins such that $R = 20$, $C = 3$. The higher the index value, the worse the performance (varied power production for fixed environmental conditions).}
\label{tab:distances_all} 
\centering
\begin{tabular}{c|c|r|r|r|r|r|r}
\hline
\hline
\multicolumn{1}{c|}{Turbine}& Temp.&\multicolumn{5}{c|}{Wind bins} &	\\\cline{3-7}
\multicolumn{1}{c|}{ID}&\multicolumn{1}{c|}{($^\circ$C)}&[5, 5.5) & [5.5, 6)&[6, 6.5)&[6.5, 7)&[7, 7.5)&\multicolumn{1}{c}{\textbf{sum}}\\

\hline\hline
\multirow{5}{*}{T01}&	15 &0.52&	0.38&	0.51&	0.32&	0.74&	\textbf{2.47}\\
&18& 0.18&	0.27&	0.28&	0.59&	1.42&	\textbf{2.73}\\
&22& 0.15	&0.28&	0.41&	0.42&	0.47&	\textbf{1.73}\\
&27&0.51	&1.29	&0.53&	0.61&	0.77&	\textbf{3.71}\\\cline{2-8}
&\textbf{sum}&\textbf{1.36}	&\textbf{2.22}	&\textbf{1.73}	&\textbf{1.94}&	\textbf{3.40}&	\textbf{10.65}\\

\hline\hline
\multirow{5}{*}{T06}&	15 & 0.29	&0.34	&0.62	&0.95	&1.55&	\textbf{3.75}\\
&18& 0.17&	0.25&	0.32&	0.29&	0.77&	\textbf{1.80}\\
&22&1.07&	0.28&	0.40&	0.39&	0.44&	\textbf{2.58}\\
&27&0.94&	0.87	&0.82&	0.65	&0.55&	\textbf{3.83}\\\cline{2-8}
&\textbf{sum}&\textbf{2.47}&	\textbf{1.74}&	\textbf{2.16}&\textbf{2.28}	&\textbf{3.31}	&\textbf{11.96}\\

\hline\hline
\multirow{5}{*}{T07}&15&0.22& 0.29& 0.37& 0.34	&1.45&	\textbf{2.67}\\
&18&0.17& 0.26 &0.34 &0.39 &0.55&\textbf{1.71}\\
&22&0.99 &0.69 &0.28& 1.25&	0.82&	\textbf{4.03}\\
&27&0.22	&0.51	&0.67	&0.42&	0.91&	\textbf{2.73}\\\cline{2-8}
&\textbf{sum}&\textbf{1.60}	&\textbf{1.75}&	\textbf{1.66}&	\textbf{2.40}&	\textbf{3.73}&\textbf{11.14}\\
\hline
\hline
\multirow{5}{*}{T11}&	15 &0.19&	0.28&	0.41&	0.36&	0.41&	\textbf{1.64}\\
&18&0.15&	0.21&	0.29&	0.40	&0.58&	\textbf{1.62}
\\
&22&0.50&	0.31&	0.30&	0.34&	0.33&	\textbf{1.78}\\
&27&0.62&	0.60&	0.76&	0.87&	0.87&	\textbf{3.73}
\\\cline{2-8}
&\textbf{sum}&\textbf{1.46}&	\textbf{1.40}&	\textbf{1.76}&	\textbf{1.97}&	\textbf{2.19}	&\textbf{8.78}\\
\hline
\hline
\end{tabular}
}
\end{table}

\section{Conclusion}
\label{sec:conclusion}

Monitoring the health of a wind turbine has to rely on data coming from sensors fitted to the components of the physical machine. One of the challenges is how to extract as much useful knowledge as one can from the available measurements. On the one hand, in many scenarios, the available input information is very limited and concerns a few variables. On the other hand, the data coming from such systems is usually noisy.

In the paper, we have presented a methodology for the visualization and quantification of wind turbine performance degradation. Our approach uses the information of the following system variables: wind speed, atmospheric temperature, and generated power to assess the machine's health. The theoretical building blocks of our proposal rely on fuzzy concepts which are labeled with symbolic terms. These fuzzy information granules allow representing and aggregating the data to capture interesting patterns. The evaluation is performed separately for various operating conditions, as we recognize their vast influence on the machine performance.

Numerical simulations concerning four wind turbines in the EDP data set showed that the turbines with ids T01 and T11 deteriorated to a smaller extent than turbines with ids T06 and T07. Both approaches, regression-based and distance index-based, support this conclusion. The regression-based method provided more distinct differentiation in the turbines ranking, meaning that the discrepancies in turbines' scores were quite large. However, we must note that this index is not expressed in natural units, but in membership degrees. In future, we will be working on concept-based models for wind turbine failure description.

\section*{Acknowledgements}\vspace{-3pt}
The project was partially funded by POB Research Centre for Artificial Intelligence and Robotics of Warsaw University of Technology within the Excellence Initiative Program - Research University (ID-UB).

\section*{Data Availability}
All experiments were conducted using a publicly available data set that can be downloaded from \url{https://opendata.edp.com/}.

\bibliographystyle{elsarticle-num-names} 
\bibliography{literature}

\begin{thebibliography}{36}
\expandafter\ifx\csname natexlab\endcsname\relax\def\natexlab#1{#1}\fi
\providecommand{\url}[1]{\texttt{#1}}
\providecommand{\href}[2]{#2}
\providecommand{\path}[1]{#1}
\providecommand{\DOIprefix}{doi:}
\providecommand{\ArXivprefix}{arXiv:}
\providecommand{\URLprefix}{URL: }
\providecommand{\Pubmedprefix}{pmid:}
\providecommand{\doi}[1]{\href{http://dx.doi.org/#1}{\path{#1}}}
\providecommand{\Pubmed}[1]{\href{pmid:#1}{\path{#1}}}
\providecommand{\bibinfo}[2]{#2}
\ifx\xfnm\relax \def\xfnm[#1]{\unskip,\space#1}\fi
\bibitem[{Jia et~al.(2018)Jia, Jin, Buzza, Di, Siegel, and Lee}]{Jia2018}
\bibinfo{author}{X.~Jia}, \bibinfo{author}{C.~Jin}, \bibinfo{author}{M.~Buzza},
  \bibinfo{author}{Y.~Di}, \bibinfo{author}{D.~Siegel},
  \bibinfo{author}{J.~Lee},
\newblock \bibinfo{title}{A deviation based assessment methodology for multiple
  machine health patterns classification and fault detection},
\newblock \bibinfo{journal}{Mechanical Systems and Signal Processing}
  \bibinfo{volume}{99} (\bibinfo{year}{2018}) \bibinfo{pages}{244--261}.
  \DOIprefix\doi{https://doi.org/10.1016/j.ymssp.2017.06.015}.
\bibitem[{Avendano-Valencia and Fassois(2017)}]{Avendano2017}
\bibinfo{author}{L.~D. Avendano-Valencia}, \bibinfo{author}{S.~D. Fassois},
\newblock \bibinfo{title}{Damage/fault diagnosis in an operating wind turbine
  under uncertainty via a vibration response gaussian mixture random
  coefficient model based framework},
\newblock \bibinfo{journal}{Mechanical Systems and Signal Processing}
  \bibinfo{volume}{91} (\bibinfo{year}{2017}) \bibinfo{pages}{326--353}.
  \DOIprefix\doi{https://doi.org/10.1016/j.ymssp.2016.11.028}.
\bibitem[{Reddy et~al.(2019)Reddy, Indragandhi, Ravi, and
  Subramaniyaswamy}]{Reddy2019}
\bibinfo{author}{A.~Reddy}, \bibinfo{author}{V.~Indragandhi},
  \bibinfo{author}{L.~Ravi}, \bibinfo{author}{V.~Subramaniyaswamy},
\newblock \bibinfo{title}{Detection of cracks and damage in wind turbine blades
  using artificial intelligence-based image analytics},
\newblock \bibinfo{journal}{Measurement} \bibinfo{volume}{147}
  (\bibinfo{year}{2019}) \bibinfo{pages}{106823}.
  \DOIprefix\doi{https://doi.org/10.1016/j.measurement.2019.07.051}.
\bibitem[{Du et~al.(2020)Du, Zhou, Jing, Peng, Wu, and Kwok}]{Du2020}
\bibinfo{author}{Y.~Du}, \bibinfo{author}{S.~Zhou}, \bibinfo{author}{X.~Jing},
  \bibinfo{author}{Y.~Peng}, \bibinfo{author}{H.~Wu},
  \bibinfo{author}{N.~Kwok},
\newblock \bibinfo{title}{Damage detection techniques for wind turbine blades:
  A review},
\newblock \bibinfo{journal}{Mechanical Systems and Signal Processing}
  \bibinfo{volume}{141} (\bibinfo{year}{2020}) \bibinfo{pages}{106445}.
  \DOIprefix\doi{https://doi.org/10.1016/j.ymssp.2019.106445}.
\bibitem[{Jacobson et~al.(2018)Jacobson, Delucchi, Cameron, and
  Mathiesen}]{Jacobson2018}
\bibinfo{author}{M.~Z. Jacobson}, \bibinfo{author}{M.~A. Delucchi},
  \bibinfo{author}{M.~A. Cameron}, \bibinfo{author}{B.~V. Mathiesen},
\newblock \bibinfo{title}{Matching demand with supply at low cost in 139
  countries among 20 world regions with 100
  sunlight (wws) for all purposes},
\newblock \bibinfo{journal}{Renewable Energy} \bibinfo{volume}{123}
  (\bibinfo{year}{2018}) \bibinfo{pages}{236--248}.
  \DOIprefix\doi{https://doi.org/10.1016/j.renene.2018.02.009}.
\bibitem[{Zhang et~al.(2020)Zhang, Wen, Liu, Jiao, Wan, and Zeng}]{Zhang2020}
\bibinfo{author}{F.~Zhang}, \bibinfo{author}{Z.~Wen}, \bibinfo{author}{D.~Liu},
  \bibinfo{author}{J.~Jiao}, \bibinfo{author}{H.~Wan},
  \bibinfo{author}{B.~Zeng},
\newblock \bibinfo{title}{Calculation and analysis of wind turbine health
  monitoring indicators based on the relationships with {SCADA} data},
\newblock \bibinfo{journal}{Applied Sciences} \bibinfo{volume}{10}
  (\bibinfo{year}{2020}). \DOIprefix\doi{https://doi.org/10.3390/app10010410}.
\bibitem[{Yang et~al.(2019)Yang, Liu, Zeng, and Xie}]{Yang2019}
\bibinfo{author}{C.~Yang}, \bibinfo{author}{J.~Liu}, \bibinfo{author}{Y.~Zeng},
  \bibinfo{author}{G.~Xie},
\newblock \bibinfo{title}{Real-time condition monitoring and fault detection of
  components based on machine-learning reconstruction model},
\newblock \bibinfo{journal}{Renewable Energy} \bibinfo{volume}{133}
  (\bibinfo{year}{2019}) \bibinfo{pages}{433--441}.
  \DOIprefix\doi{https://doi.org/10.1016/j.renene.2018.10.}
\bibitem[{{Liu} et~al.(2017){Liu}, {Shi}, {Yu}, and {Zhu}}]{Liu2017}
\bibinfo{author}{X.~{Liu}}, \bibinfo{author}{K.~{Shi}},
  \bibinfo{author}{H.~{Yu}}, \bibinfo{author}{Z.~{Zhu}},
\newblock \bibinfo{title}{Relative health index of wind turbines based on
  kernel density estimation},
\newblock in: \bibinfo{booktitle}{{IECON} 2017 - 43rd Annual Conference of the
  IEEE Industrial Electronics Society}, \bibinfo{year}{2017}, pp.
  \bibinfo{pages}{5957--5961}.
  \DOIprefix\doi{https://doi.org/10.1109/IECON.2017.8217033}.
\bibitem[{Carroll et~al.(2019)Carroll, Koukoura, McDonald, Charalambous, Weiss,
  and McArthur}]{Carroll2019}
\bibinfo{author}{J.~Carroll}, \bibinfo{author}{S.~Koukoura},
  \bibinfo{author}{A.~McDonald}, \bibinfo{author}{A.~Charalambous},
  \bibinfo{author}{S.~Weiss}, \bibinfo{author}{S.~McArthur},
\newblock \bibinfo{title}{Wind turbine gearbox failure and remaining useful
  life prediction using machine learning techniques},
\newblock \bibinfo{journal}{Wind Energy} \bibinfo{volume}{22}
  (\bibinfo{year}{2019}) \bibinfo{pages}{360--375}.
  \DOIprefix\doi{https://doi.org/10.1002/we.2290}.
\bibitem[{Wu et~al.(2019)Wu, Zhang, Yu, Jiang, and Arola}]{Wu2019}
\bibinfo{author}{R.~Wu}, \bibinfo{author}{D.~Zhang}, \bibinfo{author}{Q.~Yu},
  \bibinfo{author}{Y.~Jiang}, \bibinfo{author}{D.~Arola},
\newblock \bibinfo{title}{Health monitoring of wind turbine blades in operation
  using three-dimensional digital image correlation},
\newblock \bibinfo{journal}{Mechanical Systems and Signal Processing}
  \bibinfo{volume}{130} (\bibinfo{year}{2019}) \bibinfo{pages}{470--483}.
  \DOIprefix\doi{https://doi.org/10.1016/j.ymssp.2019.05.031}.
\bibitem[{Avendano-Valencia et~al.(2020)Avendano-Valencia, Chatzi, and
  Tcherniak}]{Avendano2020}
\bibinfo{author}{L.~D. Avendano-Valencia}, \bibinfo{author}{E.~N. Chatzi},
  \bibinfo{author}{D.~Tcherniak},
\newblock \bibinfo{title}{Gaussian process models for mitigation of operational
  variability in the structural health monitoring of wind turbines},
\newblock \bibinfo{journal}{Mechanical Systems and Signal Processing}
  \bibinfo{volume}{142} (\bibinfo{year}{2020}) \bibinfo{pages}{106686}.
  \DOIprefix\doi{https://doi.org/10.1016/j.ymssp.2020.106686}.
\bibitem[{Willis et~al.(2018)Willis, Niezrecki, Kuchma, Hines, Arwade,
  Barthelmie, DiPaola, Drane, Hansen, Inalpolat, Mack, Myers, and
  Rotea}]{Willis2018}
\bibinfo{author}{D.~Willis}, \bibinfo{author}{C.~Niezrecki},
  \bibinfo{author}{D.~Kuchma}, \bibinfo{author}{E.~Hines},
  \bibinfo{author}{S.~Arwade}, \bibinfo{author}{R.~Barthelmie},
  \bibinfo{author}{M.~DiPaola}, \bibinfo{author}{P.~Drane},
  \bibinfo{author}{C.~Hansen}, \bibinfo{author}{M.~Inalpolat},
  \bibinfo{author}{J.~Mack}, \bibinfo{author}{A.~Myers},
  \bibinfo{author}{M.~Rotea},
\newblock \bibinfo{title}{Wind energy research: State-of-the-art and future
  research directions},
\newblock \bibinfo{journal}{Renewable Energy} \bibinfo{volume}{125}
  (\bibinfo{year}{2018}) \bibinfo{pages}{133--154}.
  \DOIprefix\doi{https://doi.org/10.1016/j.renene.2018.02.049}.
\bibitem[{Yang et~al.(2018)Yang, Liu, and Jiang}]{Yang2018}
\bibinfo{author}{W.~Yang}, \bibinfo{author}{C.~Liu},
  \bibinfo{author}{D.~Jiang},
\newblock \bibinfo{title}{An unsupervised spatiotemporal graphical modeling
  approach for wind turbine condition monitoring},
\newblock \bibinfo{journal}{Renewable Energy} \bibinfo{volume}{127}
  (\bibinfo{year}{2018}) \bibinfo{pages}{230--241}.
  \DOIprefix\doi{https://doi.org/10.1016/j.renene.2018.04.059}.
\bibitem[{Chen et~al.(2021)Chen, Li, Wang, Zuo, Heyns, and
  Baggeroehr}]{Chen2021}
\bibinfo{author}{P.~Chen}, \bibinfo{author}{Y.~Li}, \bibinfo{author}{K.~Wang},
  \bibinfo{author}{M.~J. Zuo}, \bibinfo{author}{P.~S. Heyns},
  \bibinfo{author}{S.~Baggeroehr},
\newblock \bibinfo{title}{A threshold self-setting condition monitoring scheme
  for wind turbine generator bearings based on deep convolutional generative
  adversarial networks},
\newblock \bibinfo{journal}{Measurement} \bibinfo{volume}{167}
  (\bibinfo{year}{2021}) \bibinfo{pages}{108234}.
  \DOIprefix\doi{https://doi.org/10.1016/j.measurement.2020.108234}.
\bibitem[{Olauson(2018)}]{Olauson2018}
\bibinfo{author}{J.~Olauson},
\newblock \bibinfo{title}{Era5: The new champion of wind power modelling?},
\newblock \bibinfo{journal}{Renewable Energy} \bibinfo{volume}{126}
  (\bibinfo{year}{2018}) \bibinfo{pages}{322--331}.
  \DOIprefix\doi{https://doi.org/10.1016/j.renene.2018.03.056}.
\bibitem[{Ouyang et~al.(2017)Ouyang, Kusiak, and He}]{Ouyang2017}
\bibinfo{author}{T.~Ouyang}, \bibinfo{author}{A.~Kusiak},
  \bibinfo{author}{Y.~He},
\newblock \bibinfo{title}{Modeling wind-turbine power curve: A data
  partitioning and mining approach},
\newblock \bibinfo{journal}{Renewable Energy} \bibinfo{volume}{102}
  (\bibinfo{year}{2017}) \bibinfo{pages}{1--8}.
  \DOIprefix\doi{https://doi.org/10.1016/j.renene.2016.10.032}.
\bibitem[{Lei et~al.(2019)Lei, Liu, and Jiang}]{Lei2019}
\bibinfo{author}{J.~Lei}, \bibinfo{author}{C.~Liu}, \bibinfo{author}{D.~Jiang},
\newblock \bibinfo{title}{Fault diagnosis of wind turbine based on long
  short-term memory networks},
\newblock \bibinfo{journal}{Renewable Energy} \bibinfo{volume}{133}
  (\bibinfo{year}{2019}) \bibinfo{pages}{422--432}.
  \DOIprefix\doi{https://doi.org/10.1016/j.renene.2018.10.031}.
\bibitem[{Stetco et~al.(2019)Stetco, Dinmohammadi, Zhao, Robu, Flynn, Barnes,
  Keane, and Nenadic}]{Stetco2019}
\bibinfo{author}{A.~Stetco}, \bibinfo{author}{F.~Dinmohammadi},
  \bibinfo{author}{X.~Zhao}, \bibinfo{author}{V.~Robu},
  \bibinfo{author}{D.~Flynn}, \bibinfo{author}{M.~Barnes},
  \bibinfo{author}{J.~Keane}, \bibinfo{author}{G.~Nenadic},
\newblock \bibinfo{title}{Machine learning methods for wind turbine condition
  monitoring: A review},
\newblock \bibinfo{journal}{Renewable Energy} \bibinfo{volume}{133}
  (\bibinfo{year}{2019}) \bibinfo{pages}{620--635}.
  \DOIprefix\doi{https://doi.org/10.1016/j.renene.2018.10.047}.
\bibitem[{Zhan et~al.(2019)Zhan, Wang, Yi, Wang, and Xie}]{Zhan2019}
\bibinfo{author}{J.~Zhan}, \bibinfo{author}{R.~Wang}, \bibinfo{author}{L.~Yi},
  \bibinfo{author}{Y.~Wang}, \bibinfo{author}{Z.~Xie},
\newblock \bibinfo{title}{Health assessment methods for wind turbines based on
  power prediction and {Mahalanobis} distance},
\newblock \bibinfo{journal}{International Journal of Pattern Recognition and
  Artificial Intelligence} \bibinfo{volume}{33} (\bibinfo{year}{2019})
  \bibinfo{pages}{1951001}. \DOIprefix\doi{10.1142/S0218001419510017}.
  \href{http://arxiv.org/abs/https://doi.org/10.1142/S0218001419510017}{{\tt
  arXiv:https://doi.org/10.1142/S0218001419510017}}.
\bibitem[{Ren et~al.(2019)Ren, Liu, Shan, and Wang}]{Ren2019}
\bibinfo{author}{H.~Ren}, \bibinfo{author}{W.~Liu}, \bibinfo{author}{M.~Shan},
  \bibinfo{author}{X.~Wang},
\newblock \bibinfo{title}{A new wind turbine health condition monitoring method
  based on {VMD-MPE} and feature-based transfer learning},
\newblock \bibinfo{journal}{Measurement} \bibinfo{volume}{148}
  (\bibinfo{year}{2019}) \bibinfo{pages}{106906}.
  \DOIprefix\doi{https://doi.org/10.1016/j.measurement.2019.106906}.
\bibitem[{Zhang et~al.(2019)Zhang, Jiang, Li, and Li}]{Zhang2019}
\bibinfo{author}{J.~Zhang}, \bibinfo{author}{N.~Jiang},
  \bibinfo{author}{H.~Li}, \bibinfo{author}{N.~Li},
\newblock \bibinfo{title}{Online health assessment of wind turbine based on
  operational condition recognition},
\newblock \bibinfo{journal}{Transactions of the Institute of Measurement and
  Control} \bibinfo{volume}{41} (\bibinfo{year}{2019})
  \bibinfo{pages}{2970--2981}.
  \DOIprefix\doi{https://doi.org/10.1177/0142331218810070}.
\bibitem[{Tewolde et~al.(2017)Tewolde, Hoeffer, and Haardt}]{Tewolde2017}
\bibinfo{author}{S.~Tewolde}, \bibinfo{author}{R.~Hoeffer},
  \bibinfo{author}{H.~Haardt},
\newblock \bibinfo{title}{Validated model based development of damage index for
  structural health monitoring of offshore wind turbine support structures},
\newblock \bibinfo{journal}{Procedia Engineering} \bibinfo{volume}{199}
  (\bibinfo{year}{2017}) \bibinfo{pages}{3242--3247}.
  \DOIprefix\doi{https://doi.org/10.1016/j.proeng.2017.09.344},
  \bibinfo{note}{x International Conference on Structural Dynamics, EURODYN
  2017}.
\bibitem[{Koukoura(2018)}]{Koukoura2018}
\bibinfo{author}{S.~Koukoura},
\newblock \bibinfo{title}{Failure and remaining useful life prediction of wind
  turbine gearboxes},
\newblock in: \bibinfo{booktitle}{Annual Conference of the PHM Society},
  volume~\bibinfo{volume}{10}, \bibinfo{year}{2018}, pp.
  \bibinfo{pages}{5957--5961}.
  \DOIprefix\doi{https://doi.org/10.36001/phmconf.2018.v10i1.712}.
\bibitem[{Ren et~al.(2021)Ren, Liu, Shan, Wang, and Wang}]{Ren2021}
\bibinfo{author}{H.~Ren}, \bibinfo{author}{W.~Liu}, \bibinfo{author}{M.~Shan},
  \bibinfo{author}{X.~Wang}, \bibinfo{author}{Z.~Wang},
\newblock \bibinfo{title}{A novel wind turbine health condition monitoring
  method based on composite variational mode entropy and weighted distribution
  adaptation},
\newblock \bibinfo{journal}{Renewable Energy} \bibinfo{volume}{168}
  (\bibinfo{year}{2021}) \bibinfo{pages}{972--980}.
  \DOIprefix\doi{https://doi.org/10.1016/j.renene.2020.12.111}.
\bibitem[{Li et~al.(2019)Li, Li, and Zhu}]{Li2019}
\bibinfo{author}{J.~Li}, \bibinfo{author}{Q.~Li}, \bibinfo{author}{J.~Zhu},
\newblock \bibinfo{title}{Health condition assessment of wind turbine
  generators based on supervisory control and data acquisition data},
\newblock \bibinfo{journal}{{IET} Renewable Power Generation}
  \bibinfo{volume}{13} (\bibinfo{year}{2019}) \bibinfo{pages}{1343--1350}.
  \DOIprefix\doi{https://doi.org/10.1049/iet-rpg.2018.5504}.
\bibitem[{Song et~al.(2018)Song, Zhang, Jiang, and Zhu}]{Song2018}
\bibinfo{author}{Z.~Song}, \bibinfo{author}{Z.~Zhang},
  \bibinfo{author}{Y.~Jiang}, \bibinfo{author}{J.~Zhu},
\newblock \bibinfo{title}{Wind turbine health state monitoring based on a
  {Bayesian} data-driven approach},
\newblock \bibinfo{journal}{Renewable Energy} \bibinfo{volume}{125}
  (\bibinfo{year}{2018}) \bibinfo{pages}{172--181}.
  \DOIprefix\doi{https://doi.org/10.1016/j.renene.2018.02.096}.
\bibitem[{Lopez~de Calle et~al.(2019)Lopez~de Calle, Ferreiro,
  Roldan-Paraponiaris, and Ulazia}]{Lopez2019}
\bibinfo{author}{K.~Lopez~de Calle}, \bibinfo{author}{S.~Ferreiro},
  \bibinfo{author}{C.~Roldan-Paraponiaris}, \bibinfo{author}{A.~Ulazia},
\newblock \bibinfo{title}{A context-aware oil debris-based health indicator for
  wind turbine gearbox condition monitoring},
\newblock \bibinfo{journal}{Energies} \bibinfo{volume}{12}
  (\bibinfo{year}{2019}). \DOIprefix\doi{https://doi.org/10.3390/en12173373}.
\bibitem[{Tcherniak(2016)}]{Tcherniak2016}
\bibinfo{author}{D.~Tcherniak},
\newblock \bibinfo{title}{Rotor anisotropy as a blade damage indicator for wind
  turbine structural health monitoring systems},
\newblock \bibinfo{journal}{Mechanical Systems and Signal Processing}
  \bibinfo{volume}{74} (\bibinfo{year}{2016}) \bibinfo{pages}{183--198}.
  \DOIprefix\doi{https://doi.org/10.1016/j.ymssp.2015.09.038},
  \bibinfo{note}{special Issue in Honor of Professor Simon Braun}.
\bibitem[{Zhao et~al.(2018)Zhao, Liu, Hu, and Yan}]{Zhao2018}
\bibinfo{author}{H.~Zhao}, \bibinfo{author}{H.~Liu}, \bibinfo{author}{W.~Hu},
  \bibinfo{author}{X.~Yan},
\newblock \bibinfo{title}{Anomaly detection and fault analysis of wind turbine
  components based on deep learning network},
\newblock \bibinfo{journal}{Renewable Energy} \bibinfo{volume}{127}
  (\bibinfo{year}{2018}) \bibinfo{pages}{825--834}.
  \DOIprefix\doi{https://doi.org/10.1016/j.renene.2018.05.024}.
\bibitem[{Dorrego et~al.(2021)Dorrego, Rios, Hernandez-Escobedo,
  Campos-Amezcua, Iracheta, Lastres, Lopez, Verde, Hechavarria, Perea-Moreno,
  and Perea-Moreno}]{Dorrego2021}
\bibinfo{author}{J.~R. Dorrego}, \bibinfo{author}{A.~Rios},
  \bibinfo{author}{Q.~Hernandez-Escobedo}, \bibinfo{author}{R.~Campos-Amezcua},
  \bibinfo{author}{R.~Iracheta}, \bibinfo{author}{O.~Lastres},
  \bibinfo{author}{P.~Lopez}, \bibinfo{author}{A.~Verde},
  \bibinfo{author}{L.~Hechavarria}, \bibinfo{author}{M.-A. Perea-Moreno},
  \bibinfo{author}{A.-J. Perea-Moreno},
\newblock \bibinfo{title}{Theoretical and experimental analysis of aerodynamic
  noise in small wind turbines},
\newblock \bibinfo{journal}{Energies} \bibinfo{volume}{14}
  (\bibinfo{year}{2021}). \URLprefix
  \url{https://www.mdpi.com/1996-1073/14/3/727}.
  \DOIprefix\doi{10.3390/en14030727}.
\bibitem[{Kanungo et~al.(2002)Kanungo, Mount, Netanyahu, Piatko, Silverman, and
  Wu}]{Kanungo2002}
\bibinfo{author}{T.~Kanungo}, \bibinfo{author}{D.~Mount},
  \bibinfo{author}{N.~Netanyahu}, \bibinfo{author}{C.~Piatko},
  \bibinfo{author}{R.~Silverman}, \bibinfo{author}{A.~Wu},
\newblock \bibinfo{title}{An efficient k-means clustering algorithm: analysis
  and implementation},
\newblock \bibinfo{journal}{IEEE Transactions on Pattern Analysis and Machine
  Intelligence} \bibinfo{volume}{24} (\bibinfo{year}{2002})
  \bibinfo{pages}{881--892}. \DOIprefix\doi{10.1109/TPAMI.2002.1017616}.
\bibitem[{Bezdek(1981)}]{Bezdek1981}
\bibinfo{author}{J.~Bezdek}, \bibinfo{title}{Pattern Recognition with Fuzzy
  Objective Function Algorithms}, \bibinfo{publisher}{Kluwer Academic
  Publishers}, \bibinfo{year}{1981}.
  \DOIprefix\doi{https://doi.org/10.1007/978-1-4757-0450-1}.
\bibitem[{Dunn(1973)}]{Dunn1973}
\bibinfo{author}{J.~Dunn},
\newblock \bibinfo{title}{A fuzzy relative of the {ISODATA} process and its use
  in detecting compact well-separated clusters},
\newblock \bibinfo{journal}{International Journal of Cybernetics and Systems}
  \bibinfo{volume}{3} (\bibinfo{year}{1973}) \bibinfo{pages}{32--57}.
  \DOIprefix\doi{https://doi.org/10.1080/01969727308546046}.
\bibitem[{Hathaway and Bezdek(2001)}]{HathawayBezdek2001}
\bibinfo{author}{R.~Hathaway}, \bibinfo{author}{J.~Bezdek},
\newblock \bibinfo{title}{Fuzzy c-means clustering of incomplete data},
\newblock \bibinfo{journal}{{IEEE} Transactions on Systems, Man, and
  Cybernetics} \bibinfo{volume}{31} (\bibinfo{year}{2001})
  \bibinfo{pages}{735--744}.
  \DOIprefix\doi{https://doi.org/10.1109/3477.956035}.
\bibitem[{Zhang et~al.(2018)Zhang, Li, Zhang, Yu, and Lu}]{Zhang2018}
\bibinfo{author}{Y.~Zhang}, \bibinfo{author}{Z.~Li},
  \bibinfo{author}{H.~Zhang}, \bibinfo{author}{Z.~Yu}, \bibinfo{author}{T.~Lu},
\newblock \bibinfo{title}{Fuzzy c-means clustering-based mating restriction for
  multiobjective optimization},
\newblock \bibinfo{journal}{International Journal of Machine Learning and
  Cybernetics} \bibinfo{volume}{9} (\bibinfo{year}{2018})
  \bibinfo{pages}{1609--1621}.
  \DOIprefix\doi{https://doi.org/10.1007/s13042-017-0668-6}.
\bibitem[{Pimentel and de~Souza(2018)}]{Pimentel2018}
\bibinfo{author}{B.~A. Pimentel}, \bibinfo{author}{R.~M. C.~R. de~Souza},
\newblock \bibinfo{title}{A generalized multivariate approach for possibilistic
  fuzzy c-means clustering},
\newblock \bibinfo{journal}{International Journal of Uncertainty, Fuzziness and
  Knowledge-Based Systems} \bibinfo{volume}{26} (\bibinfo{year}{2018})
  \bibinfo{pages}{893--916}.
  \DOIprefix\doi{https://doi.org/10.1142/S021848851850040X}.

\end{thebibliography}


%




%
%
%
\end{document}